\documentclass{tcibook}
\usepackage{fancyhea}
\usepackage{work}
\usepackage{bm}       %    enables bold math symbols  e.g.  \bm{\gamma}
\usepackage{graphicx}
\usepackage{hyperref}      % hypertext links %%ARXIV
\usepackage{picture}
\usepackage{lineno}

\newcommand{\nc}{\newcommand}  

\def\etal{{\it et al.}}

\def\Acknowledgements{\bigskip  \bigskip \begin{center} \begin{large}
             \bf ACKNOWLEDGEMENTS \end{large}\end{center}}

\def\beq{\begin{equation}}
\def\eeq#1{\label{#1}\end{equation}}
\def\eeqn{\end{equation}}

\newenvironment{Eqnarray}%
   {\arraycolsep 0.14em\begin{eqnarray}}{\end{eqnarray}}
\def\beqa{\begin{Eqnarray}}
\def\eeqa#1{\label{#1}\end{Eqnarray}}
\def\eeqan{\end{Eqnarray}}

\def\leqn#1{(\ref{#1})}

\nc{\ra}{\rightarrow}  
\nc{\slsh}{\slash\hspace*{-0.22cm}}
\def\Re{{\cal R \mskip-4mu \lower.1ex \hbox{\it e}\,}}
\def\Im{{\cal I \mskip-5mu \lower.1ex \hbox{\it m}\,}}

\nc{\vev}[1]{ \left\langle {#1} \right\rangle }
\nc{\bra}[1]{ \langle {#1} | }
\nc{\ket}[1]{ | {#1} \rangle }
\nc{\fb}{\,{\rm fb}^{-1}}
\nc{\ev}{{\rm eV}}
\nc{\kev}{{\rm keV}}
\nc{\Mev}{{\rm MeV}}
\nc{\gev}{{\rm GeV}}
\nc{\tev}{{\rm TeV}}
\nc{\mev}{{\rm MeV}}

\def\L{{\cal L}}

\def\del{\partial}
\def\Dslash{\not{\hbox{\kern-4pt $D$}}}
\def\dslash{\not{\hbox{\kern-2pt $\del$}}}
\def\pslash{\not{\hbox{\kern-2pt $p$}}}
\def\ETmiss{ \not{\hbox{\kern-4pt $E$}}_T }

\def\tr{\ensuremath{\mbox{\rm tr}}\xspace}

\def\ee{e^+e^-}

\def\msb{{\bar{\ssstyle M \kern -1pt S}}}

\newcommand{\cms}{cm$^{-2}$s$^{-1}$}

\usepackage{enumitem}

\begin{document}
%\linenumbers

\def\bibname{References}
\bibliographystyle{plain}

\raggedbottom

\pagenumbering{roman}

\parindent=0pt
\parskip=8pt
\setlength{\evensidemargin}{0pt}
\setlength{\oddsidemargin}{0pt}
\setlength{\marginparsep}{0.0in}
\setlength{\marginparwidth}{0.0in}
\marginparpush=0pt

% The content begins here

\pagenumbering{arabic}

\renewcommand{\chapname}{chap:intro_}
\renewcommand{\chapterdir}{.}
\renewcommand{\arraystretch}{1.25}
\addtolength{\arraycolsep}{-3pt}

\pagenumbering{roman}

\thispagestyle{empty}
\begin{centering}
\vfill

{\Huge\bf Planning the Future of U.S. Particle Physics}

{\Large \bf Report of the 2013 Community Summer Study}

\vfill

{\Huge \bf Chapter 3:  Energy Frontier}

\vspace*{2.0cm}
{\Large \bf Conveners: R. Brock and M. E. Peskin}
\pagenumbering{roman}

\vfill

{\large  Study Conveners: M. Bardeen, W. Barletta, L.~A.~T.~Bauerdick, R. Brock,
D.~Cronin-Hennessy, M.~Demarteau, M.~Dine, J.~L. Feng, M. Gilchriese,
S. Gottlieb, J.~L.~Hewett, R. Lipton, H.~Nicholson, M.~E. Peskin,
S. Ritz, I.~Shipsey, H. Weerts}\\
\vspace{1cm}

{\large Division of Particles and Fields Officers in 2013:
J.~L. Rosner (chair), 
I. Shipsey (chair-elect), 
N. Hadley (vice-chair),
P. Ramond (past chair)}\\
\vspace{1cm}

{\large Editorial Committee:
R.~H. Bernstein,
N. Graf,
P. McBride,
M.~E. Peskin,
J.~L. Rosner,
N.~Varelas,
K. Yurkewicz}

\vfill

\end{centering}

\newpage
\mbox{\null}

\vspace{3.0cm}

{\Large \bf Authors of Chapter 3:}

\vspace{2.0cm}
 {\bf  R. Brock, M. E. Peskin}, K. Agashe, M. Artuso, J. Campbell, S. Dawson, R. Erbacher, C. Gerber, Y.~Gershtein, A. Gritsan, K. Hatakeyama, J. Huston, A. Kotwal,  H. Logan, M. Luty, K. Melnikov, M.~Narain, M. Papucci, F. Petriello, S. Prell,  J. Qian, R. Schwienhorst, C. Tully, R. Van Kooten, D. Wackeroth, L.~Wang, D. Whiteson

 \tableofcontents

\newpage

\mbox{\null}

\newpage

\pagenumbering{arabic}

%%%%%%%%%%%%%%%%%%%%%%%%%%%%%%%%%%%%%%%%%%%%%%%%%%%
%%%%%%%%%%%%%%%%%%%%%%%%%%%%%%%%%%%%%%%%%%%%%%%%%%%
%%%     All of your files should be in a subdirectory.  Here the
%%%     subdirectory is called Magnetism  .   The title of your
%%%     report should be   wgreport.tex in that subdirectory.  Input
%%%     that file here
%%%%%%%%%%%%%%%%%%%%%%%%%%%%%%%%%%%%%%%%%%%%%%%%%%%%
%%%%%%%%%%%%%%%%%%%%%%%%%%%%%%%%%%%%%%%%%%%%%%%%%%%

%\input Magnetism/wgreport.tex 
% !TEX root = ../EF_driver_commenting.tex
%%%%%% EF Chapter  %%%%%%%%%%%%%%%%

% text mode macro
\def\ifb{fb$^{-1}$}
\def\MSbar{$\overline{MS}$}
\def\ssteff{\sin^2\theta_{eff}}
% math mode macros
\def\tr { \mbox{tr} }
\def\ETmiss{ \not{\hbox{\kern-4pt $E$}}_T }
\def\dzero{ \not{\hbox{\kern-2.6pt \mbox{O}}} }

\setcounter{chapter}{2}

\chapter{Energy Frontier} 
\label{chap:ef}

%%%%%%%%%%%%%%%%%%%%%%%%%%%%%%%%%%%%%%%%%%%%%%%%%%%%%%%%%%%
%%%%%%%%%%%%%%%%%%%%%%%%%%%%%%%%%%%%%%%%%%%%%%%%%%%%%%%%%%%
%%%%%%%%%%%%%%%%%%%%%%%%%%%%%%%%%%%%%%%%%%%%%%%%%%%%%%%%%%%
%%%%%%%%%%%%%%%%%%%%%%%%%%%%%%%%%%%%%%%%%%%%%%%%%%%%%%%%%%%
\begin{center}\begin{boldmath}

% list of HSPAW authors

%\hyphenpenalty 10000

\begin{center}

\begin{large} {\bf Conveners: R. Brock, M. E. Peskin} \end{large}

K.~Agashe, M.~Artuso, J.~Campbell,  S.~Dawson, R.~Erbacher, C.~Gerber,
Y.~Gershtein, A.~Gritsan, 
K.~Hatakeyama, J.~Huston, 
A.~Kotwal, 
H.~Logan, M.~Luty, K.~Melnikov, M.~Narain, M.~Papucci, F.~Petriello, S.~Prell,  J.~Qian, R.~Schwienhorst,
C.~Tully,
R.~Van Kooten, D.~Wackeroth, L.~Wang, D.~Whiteson

\end{center}

%\hyphenpenalty 1000

%Conveners are also listed separately in authorlist.tex

\end{boldmath}\end{center}

%%%%%%%%%%%%%%%%%%%%%%%%%%%%%%%%%%%%%%%%%%%%%%%%%%%%%%%%%%%
%%%%%%%%%%%%%%%%%%%%%%%%%%%%%%%%%%%%%%%%%%%%%%%%%%%%%%%%%%%
%%%%%%%%%%%%%%%%%%%%%%%%%%%%%%%%%%%%%%%%%%%%%%%%%%%%%%%%%%%
%%%%%%%%%%%%%%%%%%%%%%%%%%%%%%%%%%%%%%%%%%%%%%%%%%%%%%%%%%%

\section{Introduction} \label{sec:EF-intro} 

The original goal of elementary particle physics was to understand the nature of the subnuclear strong, electromagnetic, and weak forces.  In the late 1960's and early 1970's, specific models for these forces were proposed in the form of Yang-Mills gauge theories, giving a beautiful explanation of all three interactions from a unified point of view.  Together, these theories became known as the ``Standard Model.''   Today, we have a great deal of confidence that describing fundamental forces using the gauge principle is correct. Through precision experiments involving $W$ and $Z$ bosons carried out over the past twenty-five years, we have tested the Standard Model stringently, and the theory has passed every test.  The most recent such experiments included the search for the Higgs boson,  required in the Standard Model to generate quark, lepton, and vector boson masses. A year ago, the ATLAS and CMS experiments at the Large Hadron Collider discovered a candidate for this particle which, at the current level of our measurements, has all of the properties predicted in the Standard Model.

This is an historic  level of success for theory and experiment: This economical model predicted the existence of fundamental fields, their dynamics, a scalar field responsible for the breaking of a gauge symmetry, and interactions among the particles  with precision unmatched in the history of science. It all seems to have come true with remarkable accuracy.    And yet, we find the result still unsatisfying.  It is typically true in science that revolutionary changes in our understanding lead to a new set of vexing questions.  The success of the Standard Model is no different.  Though we have answered many questions about the structure of elementary particles, we have a new set of questions about the structure of the Standard Model itself.   The discovery of the Higgs boson sharpens these issues and makes them even more mysterious.

There are many phenomena in nature that obviously lie outside of the Standard Model.  
\begin{itemize}
\item We now know that 85\% of the matter in the universe is {\it dark matter} --- neutral, weakly interacting matter composed of one or more species not contained in the Standard Model.    
\item The excess of baryons over antibaryons in the universe 
is not explained by the Standard Model.  Even though the Standard Model contains
all of the necessary ingredients to generate baryon number in the early universe, including baryon number violation, CP violation, and a phase transition in cosmic history, 
it predicts a baryon-antibaryon asymmetry that  is too small by ten orders of magnitude. 
\item The quantum numbers of the quarks and leptons under the Standard Model gauge symmetry $SU(3)\times SU(2)\times U(1)$ strongly suggests that these symmetry groups are unified into a larger {\it grand unification} group $SU(5)$ or $SO(10)$; however,  the results of precision measurements
 of the strengths of the gauge couplings is inconsistent with this hypothesis.  
\item The Standard Model cannot account for neutrino masses without the addition of some new particles.  
\item Further, the pattern of  weak interaction mixing among neutrinos is completely different from that observed for quarks. 
\item The Standard Model does not include the force of gravity or the small but nonzero energy  in empty space that gives rise to {\it dark energy}. 
\end{itemize}
In addition, there is a major theoretical puzzle with the Standard Model. If the Higgs boson is an elementary scalar particle,  its mass is sensitive to the masses of any heavier particle to which it couples. It appears to require a cancellation of one part in $10^{32}$ to explain why the Higgs boson mass is smaller than the Planck mass.

The discovery of the Higgs boson has changed our viewpoint in how we address these unexplained phenomena and theoretical questions.  This is true for three reasons.  \\ \\
{\bf First, the Higgs boson completes the particle spectrum of the Standard Model.}  We have now discovered all of the Standard Model particles and have measured many of their properties.  It is clear now exactly what the model does {\it not } explain.  We have entered a new era in which the verification of the Standard Model takes second place to a search for new, unknown forces and interactions.

{\bf Second, one of the key mysteries concerns the Higgs boson itself.} The Higgs boson was predicted as a direct consequence of the simplest model of the generation of mass for quarks, leptons, and gauge bosons. For a long time, many particle physicists have expressed discomfort with this model.
Now the prediction has become a reality. We have to grapple with it and understand why nature chooses a particle with these properties to do its work. 

{\bf Third, the Higgs boson itself gives us a new experimental approach.} Within the Standard Model, all
properties of the Higgs boson are precisely predicted from the value of the Higgs mass.  But, as soon as we step outside the Standard Model, the properties of the Higgs boson are hardly constrained by theory.   It is compelling to tug on this particle until the Standard Model breaks.   We need to apply to the Higgs boson  the same scrutiny that we have applied in previous decades to hadron structure, heavy quark system, the $W$ and $Z$ bosons, and top quark.  Each study was done at the Energy Frontier machines of its day. This fruitful experimental approach has acquired a new, promising target.

For exploration of the unknown regions outside the Standard Model, we are encouraged that very powerful experimental tools will be put into play.   In the next ten years, the Large Hadron Collider (LHC) at CERN is expected to almost double its energy and to  increase its total  event sample, or ``integrated luminosity,'' from the current 25~\ifb\ to 400~\ifb.
This new capability will put to the test many models that predict new physics beyond the Standard Model and 
address the unexplained  phenomena listed earlier in this section.
 In the decade after that, the LHC should increase its data set by a further factor, up to 3000~\ifb.
Particle physicists have proposed
lepton colliders and higher energy hadron colliders with capabilities beyond those of 
the LHC.  The mysteries associated with the Higgs boson call for new particles and forces at the TeV energy scale or the attometer distance scale.  We now have  before us tools
 for a thorough exploration of this region of masses and distances.  
This is a compelling program. The purpose of this report is to describe it in detail. 

The structure of this summary report is as follows:   In Section \ref{sec:EF-motiv}, we present the arguments for new fundamental interactions at the TeV energy scale and the experimental program at colliders that these arguments motivate. Section \ref{sec:EForg}  describes  the organization of the Energy Frontier study.  In Sections~\ref{sec:higgs}--\ref{sec:flavor}, we review in a more specific way the physics issues of collider experiments at the TeV energy scale.  We consider in turn the prospects for exploration of new physics through studies of the Higgs boson, the $W$ and $Z$ bosons, quantum chromodynamics (QCD), the top quark, and searches for and study of new particles. We present the questions that need to be answered and the methodologies to attack these questions.  In Section~\ref{sec:cases}, we present the capabilities of current and proposed colliders in relation to these physics goals. Finally, in Section~\ref{sec:disc-stor}, 
we trace out the implications of two possible scenarios for the discovery of new physics at the LHC.
This gives an orthogonal way of appreciating the contributions that might be made by 
 proposed accelerators.  Section~\ref{sec:Messages} gives the summary statements from each of the 
working groups, and 
 Section~\ref{sec:EF-Concl} gives our overall conclusions.

\section{Importance of the TeV Scale}  \label{sec:EF-motiv}

We have listed a number of motivations for new fundamental interactions beyond the Standard Model (SM).   Where will we find them?

Explanations for baryogenesis, higher unification, and dark energy span a bewildering range of mass and distance scales.  However, many of the questions we have listed in the previous section relate specifically to the energy scale of hundreds to thousands of GeV that we are exploring today at the LHC.   We consider it imperative to understand particles and forces at this ``TeV scale'' thoroughly, using all of the tools at our disposal.  In this section, we will discuss the importance of this regime of energies and short distances.

There is a sharp boundary at which our well-founded knowledge of the fundamental elementary particle interactions runs out. This is related to two different faces that the SM presents, which stand on very different theoretical foundations.  On one side are the Yang-Mills gauge interactions. On the other side are the interactions of the Higgs field.  The Yang-Mills interactions of quarks, leptons, and vector bosons are tightly determined by their quantum numbers and the strength of the coupling constants of the $SU(3)\times SU(2)\times U(1)$ vector bosons.  Precision tests of the SM confirm the structure of these interactions to impressive accuracy~\cite{LEPEW}.  There is little doubt that the SM is a correct representation of nature at the 
energies we have currently explored.

On the other hand, the interactions of the SM fermions with the Higgs field, and the dynamics of the Higgs field itself,  are essentially unconstrained and conceptually cumbersome.   The SM Lagrangian is constructed by writing down the most general terms allowed by gauge symmetry and renormalizability. The resulting  potential term contains much of what is perplexing about the SM:
\beq    V=  \mu^2\Phi^\dagger \Phi+\lambda(\Phi^\dagger\Phi)^2 + \sum_{f,f'}  \left[ g_{\Phi ff'}\bar{f}_{L}f'_{R}\Phi + h.c.\right].
\eeq{higgspot} 
%
%\left[ \frac{g_{Hff}v}{\sqrt{2}}\bar{f}_{L}f_{R} + \frac{g_{hff}}{\sqrt{2}}\bar{f}_{L}f_{R} + h.c.
%
%
Here $\Phi$ is a spin 0 field and $\lambda$, $g_{\Phi ff'}$,  and $\mu$ are parameters.  The first two terms give the potential energy of the $\Phi$ field. When $\mu^2<0$, the state of lowest energy or vacuum state is obtained by  shifting the $\Phi$ field to a nonzero overall value. The shift is written  $\Phi^0 \to (v + h)/\sqrt{2}$, where  $v= \sqrt{-\mu^2/\lambda}$.  The quantity $v$ is the Higgs field vacuum value,
which gives mass to all quarks, leptons, and gauge bosons.  The field $h(x)$ is the quantum Higgs field, whose quanta are the 
Higgs bosons.  In this report, we will refer to the process in  which the $\Phi$ field takes on a nonzero value at every point
in space as ``Higgs condensation''.  

 At the same time, we diagonalize the matrix $g_{\Phi ff'}$. The last term in brackets then becomes 
\beq
 V(ff) = \sum_f \left[ \frac{g_{Hff}v}{\sqrt{2}}\bar{f}_{L}f_{R} + \frac{g_{Hff}}{\sqrt{2}}\bar{f}_{L}f_{R}h + h.c. \right].
\eeq{ff}
The first term in the brackets is a mass term for each type of fermion, quark or lepton.  
The second term gives the corresponding interaction, or  Yukawa coupling,  of each fermion with the Higgs boson.

Every term in \leqn{higgspot} and \leqn{ff} is of critical importance and each presents special challenges to interpretation and measurement. The  fermion masses are given by  $m_f = {g_{Hff}v}/{\sqrt{2}}$, proportional
to the couplings to the Higgs particle.  The pattern of the fermion masses is totally unconstrained.  The eigenvalues of the matrix $g_{Hff}$  contains the origin of  mass and mixings among 
quarks and leptons and of CP violation in the weak interactions. Many
of these masses and mixing angles are well measured, but there is no theory that
 explains their origin and structure.  

As to the couplings of the Higgs boson itself {(the first two terms of \leqn{higgspot})}, the picture given in the Standard Model is just one choice among many possibilities. There may be additional Higgs bosons and additional particles of other types forming a larger ``Higgs sector.''  We have almost no information about these particles except that their  masses are probably larger than the mass of the known Higgs boson at 125~GeV,

\subsection{The mystery of Higgs field symmetry breaking} \label{sec:EF-TeV}

The issues raised in Section~\ref{sec:EF-intro} are brought to a focus by a single underlying question:   In order for any SM quark, lepton, or vector boson to obtain mass, the Higgs field must condense and acquire a 
nonzero value throughout the universe.   Why does this happen?

The SM itself provides no help with this question.  It states only that symmetry 
breaking occurs if $\mu^2 < 0$, which, as a physics explanation,
is completely empty. Potentials of the form of \leqn{higgspot} appear in many condensed matter systems, including superconductors, magnets, and binary alloys.  In those systems, it is possible to compute the parameters of the potential from the underlying details of atomic structure and explain why $\mu^2 < 0$.  For the SM, if there is an underlying dynamics, its form is unknown.   Attempts to compute $\mu^2$ within the SM, even to determine its sign, give disastrous results. The answers for $\mu^2$ depend quadratically on the values of large, unknown mass scales, with competing contributions of opposite sign.

There are models, generalizing the SM, in which $\mu^2$ can be computed.  
However, these are not simple extensions of the SM.  The barrier to creating 
such a model is that the quadratic dependence 
on unknown scale parameters at very high energy must be removed.    
However, this dependence is a generic property of models with fundamental
 scalar fields, associated with the fact that the radiative corrections to the
 scalar field mass are quadratically divergent.  Cures for this problem
 require that the Higgs particle be non-generic in some important way: 
Either it is a composite particle  or it  is related by a symmetry to a fermion
 or a vector boson.  Symmetries of these types can be included consistently 
only by profound extension of the structure of space-time, to supersymmetry 
in the fermion case or higher dimensions of space in the vector boson case.

It is remarkable that, in each of these classes of models, easily identified radiative corrections give contributions to $\mu^2$ with a negative sign, predicting the instability of the Higgs field to condensation~\cite{refsformu}. In all of the models mentioned in the previous paragraph, these 
contributions come from quantum corrections due to partners of the top quark that are 
required by the postulated new interactions.

The idea that the condensation of the Higgs field has a definite mechanical explanation from quantum physics thus has major implications. It requires a new set of particles at the TeV mass scale.  The examples above include exotic partners of the top quark that are likely to  be produced at the LHC.  A TeV particle spectrum can also supply explanations for other issues that require physics beyond the SM.  TeV particle spectra typically contain a massive neutral particle that can be absolutely stable and thus a candidate for the particle of dark matter~\cite{SUSYDM}.  New couplings among the TeV particles potentially provide new sources of CP violation, offering mechanisms for creating the matter-antimatter asymmetry of the universe. Corrections to the SM coupling constants from the new particles can correct the evolution of the SM couplings, allowing the three SM gauge interactions to unify at very short distances~\cite{unif}.

Most importantly, if the explanation for Higgs condensation changes our view of the SM itself --- by making SM particles composite or by enlarging the structure of space-time --- these changes must be taken into account in any explanation of phenomena that occur at still smaller distances scales, including the generation of neutrino  masses, the generation of flavor mixing among quarks and leptons, and the unification of the particle physics interactions with gravity.

In short: mechanisms that shed light on the physics behind the otherwise mysterious potential in \leqn{higgspot} are needed to directly address the major experimental anomalies that require physics
beyond the SM.

\subsection{Naturalness} \label{sec:Natural}

To test models that postulate new  particles, we must find and characterize those particles. 
To do this, it is helpful to know to what high energy we must probe.  Unfortunately, there is no 
crisp answer to this question.  We  have only a hint 
from the principle of ``naturalness.''

Naturalness is the statement that new particles that generate the $\mu^2$ term in the Higgs potential \leqn{higgspot}  must have masses at the scale of $\mu^2$ itself, 
\beq    \mu^2 \sim   (100~\mbox{GeV})^2   \ . \eeq{muest} 
Taken most naively, naturalness implies that new particles associated with the Higgs potential should have been found in the 1990's at the experiments at LEP and the Tevatron.  Today, the LHC experiments have carried out much deeper searches for these particles.   How much further must we go?

One approach to naturalness looks more critically at the radiative corrections to the $\mu^2$ parameter in the SM.   The first-order corrections due to the top quark, the $W$ and $Z$ bosons, and the Higgs boson itself are \beq   
 \delta \mu^2 = - {3g_{Htt}^2\over 8\pi^2}\Lambda^2 + {3      \alpha_w(3+\tan^2\theta_W)\over 4\pi} \Lambda^2 + {\lambda\over      8\pi^2}\Lambda^2 \ , 
\eeq{muformula}
 where $g_{Htt}$ is the same Yukawa coupling as in \leqn{ff}, $\alpha_w$ and $\lambda$ are the couplings of the $W$, $Z$, and Higgs bosons,  and $\theta_W$ is the weak mixing angle.  All three terms are divergent and proportional to $\Lambda^2$, where $\Lambda$ is a mass scale at which the SM is replaced by a more complete underlying theory.  Contributions from new particles add to (or subtract from)  this expression. If the sum of SM and new particle contributions is to give a well-defined result for $\mu^2$, the new terms must cancel the 
 dependence on $\Lambda$ in \leqn{muformula}.   If we allow the new contributions to cancel the SM ones over many decimal places, $\Lambda$ can be  arbitrarily high. However, this might be considered ``unnatural.''  If we assume that at most one significant figure is cancelled, we obtain interesting limits on top quark, $W$ boson, and Higgs boson partners at roughly 2~TeV.

Another approach looks into the computation of $\mu^2$ in specific models~\cite{NaturalSUSY}.  In most supersymmetry (SUSY) models, the parameter called $\mu_{SUSY}$ --- the Higgsino mass term --- contributes to the Higgs parameter $\mu^2$ at the tree level.  Forbidding cancellations beyond one significant figure gives limit on the SUSY parameter, $\mu_{SUSY} < 200$~GeV.  This is a strong upper bound on the mass of the supersymmetric partner of the Higgs boson, a particle that will be difficult to discover at the LHC.  The supersymmetric partners of the top quark ($\widetilde t$) and the gluon ($\widetilde g$) contribute to the Higgs potential in one-loop and two-loop order, respectively.  The corresponding naturalness bounds for one-significant-figure cancellations
are 
\beq   
  m(\widetilde t) < 1~\mbox{TeV}  \quad \mbox{and}  \quad  m(\widetilde g) < 2~\mbox{TeV} \ . 
\eeq{stopglubounds} 

In models in which the Higgs boson is a composite Nambu-Goldstone boson, the formula for the radiative correction to $\mu^2$ from a new fermionic partner $T$ of the top quark has the form \beq   
  \delta \mu^2  =   C  {3\lambda_t^2\over 8\pi^2} m(T)^2 \ , 
\eeq{muTbound} 
where $C$ is a model-dependent constant of order 1.  This gives a bound for one-significant-figure cancellation
\beq        
     m(T) <   2~\mbox{TeV} \ . 
\eeq{Tbound}

In all cases, we might have stronger cancellations in the expressions for $\mu^2$. 
These cancellations might eventually find some physics explanation.   However, 
each factor of 10 in mass above the bounds quoted requires cancellations of another
 {\it two} significant figures.  Even such an imprecise criterion as naturalness
 probably limits top quark partners to lie below about 10~TeV.

However unsatisfactory these naturalness estimates might be, our interest in these estimates remains very strong.   Higgs condensation is the mechanism that generates the whole spectrum of masses of the SM quarks, leptons, and vector bosons.  Can it be just an accident?  If not, there {\it must} be a spectrum of new particles at the TeV scale.   Even if we cannot predict the value of this scale incisively, the importance of mass scale is clear.  We must find these new states.

\subsection{The mystery of dark matter} \label{sec:dark}

Independent of the naturalness argument, there is another argument for new particles at
the TeV mass scale.  The Standard Model does not account for the dark matter which 
makes up 85\% of the total matter content of the universe.  Among the many explanations for dark 
matter, there is one that is particularly compelling.  This is the model of dark matter as composed
of a neutral, weakly-interacting, massive particle (WIMP) that was produced in the hot conditions of the
early universe.  The WIMP model of dark matter is discussed in full detail in the Cosmic Frontier 
report~\cite{CFreport}.  What is interesting here is one implication of the model:  In order to obtain the
observed density of dark matter, the energy scale of the interactions of dark matter must be close to 
1~TeV.  The TeV mass scale arises as a combination of astrophysical parameters with no obvious
relation to the Higgs potential.  This could be a coincidence, but it might be
 a suggestion that models for the Higgs potential also solve the dark matter problem.

Theoretical models with WIMPs typically contain many particles with TeV scale masses.  These 
particles share a common quantum number.  They decay to the lightest particle carrying this quantum 
number, which is then stable for the lifetime of the universe.  Whether or not this new spectrum of 
particles is connected to the Higgs problem, it is important use colliders to search for the 
new particles required by these models.

\subsection{Summary}

The ideas reviewed in this section
 predict a spectrum of new particles at the TeV mass scale.  Those particles should be
discoverable in experiments at the LHC and  planned future accelerators. These experiments will provide  the crucial tests of those ideas. Furthermore, if such particles are discovered, they can be studied in detail in collider experiments to determine their properties and  to establish new fundamental laws of nature.

A research program in pursuit of new particles with TeV masses consists of three threads:
\begin{enumerate}
\item  We must study the Higgs boson itself in as much detail as possible, searching for signs of a larger Higgs sector and the effects of new heavy particles.
\item We must search for the imprint of the Higgs boson  and its possible partners on the couplings of
the $W$ and $Z$ bosons and the top quark.
\item We must search directly for new particles with TeV masses that can address  important problems in fundamental physics.
\end{enumerate}
To the extent that the naturalness or dark matter arguments above are a guide, all three approaches will be accessible at high-energy collider experiments in the near future.   In the next section, we will describe the tools that we have available for that search.

\section{Organization of the Energy Frontier study}
\label{sec:EForg}

We divided the study of the TeV energy scale thematically, in terms of probes of this scale using different particles and interactions.  The participants in the study were asked to enunciate and 
compare the capabilities of a variety of operating and planned accelerators for each experimental
approach.  The results summarized here constitute the efforts of hundreds of physicists who 
worked through the winter and spring of 2013 within six working groups. The leaders of these groups are co-authors of this report.

\subsection{Working groups for the study of the Energy Frontier}

 The Energy Frontier working groups covered the following topics:
\begin{enumerate}
\item The Higgs Boson
\item Electroweak Interactions
\item Quantum Chromodynamics and the Strong Force
\item Understanding the Top Quark
\item The Path Beyond the Standard Model - New Particles, Forces, and Dimensions
\item Flavor Mixing and CP Violation at High Energy
\end{enumerate}
Highlights of each group's work  are presented in this order in the following six sections.
We follow with the scientific cases to be made for each possible accelerator organized around each physics group's conclusions for that facility.  We summarize the most important conclusions of each group 
in Section~\ref{sec:Messages}.

\subsection{Accelerators for the Study of the Energy Frontier} \label{sec:accel}

Specific estimates of the capabilities of the methods that we discussed are made in the context of proposed accelerator programs discussed at Snowmass. We provide here a brief orientation to these programs.  Energies refer to the center-of-mass energy of colliding beam experiments. For details on the design and current status of these proposals, see the Capabilities Frontier working group report \cite{CapabilitiesWG}.

The baseline for our study is the Large Hadron Collider (LHC), the $pp$ collider now operating at CERN. The 
most recent LHC schedule calls for 75-100~\ifb\ to be collected in a run
 starting in 2015 using  the current detectors with minor upgrades. Following 
a shutdown  in approximately 2018, the Phase 1 detector upgrades 
will be installed and running will resume at a
projected instantaneous luminosity of  $2\times 10^{34}$cm$^{-2}$s$^{-1}$.  This stage would produce another
300~\ifb\ of data. In approximately 2023, after another shutdown,
 the luminosity is expected to increase to $5\times 10^{34}$cm$^{-2}$s$^{-1}.$  In the Snowmass study, we compared the current results from the LHC, at 7-8~TeV with an integrated luminosity of 20~\ifb, to future data samples at 14 TeV with 300~\ifb\ and with 3000~\ifb.  We refer to the latter program as the high-luminosity LHC or HL-LHC.  The projected evolution of the LHC program is described in~\cite{LHCfuture}.

Our study considered higher energy $pp$ colliders, with energy 33 TeV and 100 TeV.  Unless it is indicated otherwise, the event sample assumed is 3000~\ifb.  A  high-energy upgrade of the LHC to 33 TeV (HE-LHC) is discussed in \cite{HELHC}.  Colliders of 100 TeV energy are described in \cite{VHELHC,VLHC}.  In the following we will refer to such a collider generically as VLHC.

Our study considered $\ee$ linear colliders, both the International Linear Collider (ILC), covering the energy range 90 GeV--1000 GeV and the Compact Linear Collider (CLIC), covering the energy range 350~GeV--3000~GeV. The ILC is described in \cite{ILCSnowmass} and in its technical design report~\cite{ILCTDR}.  The TDR/CDR luminosity samples are 1000~\ifb\ at 1~TeV and scale linearly with energy. Luminosity upgrades of the baseline ILC using strategies outlined in the TDR, to 2500~\ifb\ at 1~TeV and similar enhancements at other energies, with long running periods, are described in \cite{ILCup}.  CLIC is described in \cite{CLICSnowmass} and in its Conceptual Design Report \cite{CLICCDR}.

Our study considered $\mu^+\mu^-$ colliders operating over a range from 125~GeV to 3000~GeV.   The luminosity samples assumed were similar to those for linear $\ee$ colliders.   The technology of the muon collider is described in \cite{MuonC,MuonHiggs}.

Our study considered a circular $\ee$ collider in a large (80-100 km) tunnel.   Accelerator parameters for such a machine are described in \cite{TLEPSnowmass} in the context of one such  proposal, TLEP, for a large tunnel near CERN.   In principle, accelerator techniques invented for super-$B$-factories can produce very high luminosities, in excess of 10$^{36}$cm$^{-2}$s$^{-1}$ at 90 GeV and 10$^{35}$cm$^{-2}$s$^{-1}$ at 250 GeV,
 summed over 4 detectors. However, there is as yet no complete accelerator design.  The luminosities are lower, and/or the power consumption rises dramatically, at 350~GeV and above.  In the following, we will refer to such a collider as TLEP (wherever it might be built). We will assume the above luminosities and operation with four detectors.

Two more types of accelerators received more limited attention from our study. Linear $\ee$ colliders can be converted to photon-photon colliders, with roughly 80\% of the energy and similar luminosity, by backscattering laser light from the electron beams.  Proposals for photon-photon colliders are described in \cite{Sapphire,Fermiphph}. Colliding the LHC beam with an $e^-$ or $e^+$  beam from a linear accelerator offers the opportunity of high energy $ep$ collisions.  This has been studied for a facility at CERN called LHeC, described in \cite{LHeCrpt}.

%\subsection{Organization of the Energy Frontier}
%
%\chipq{I thought that the actual groups should be referred to somewhere. \\} 
%\mq{Excuse me, but what do you have in mind? Put the references to the working group reports here?}
%
%\chipq{Yikes. I had a bulleted list of the working groups and it's gone! Wait. I did do it and it's above in 1.3.1. I'll leave this little stream of consciousness in here to answer your question. This whole 1.3.3 section goes away.}

% %

\line(1,0){2.5in}
\vspace{-12pt}
\section{The Higgs Boson} \label{sec:higgs} %
\vspace{-12pt}\line(1,0){2.5in}
\vspace{12pt}

We begin with the study of the Higgs boson itself.  In this section, we will refer to the new boson with mass 125~GeV as ``the Higgs boson,'' while recognizing that its properties could well be very different from the simplest expectations.

We have already emphasized that the study of the Higgs boson gives a completely new avenue along which to probe the physics  of the TeV scale.  The picture of the Higgs boson given by the SM is precise.  All properties of the Higgs boson can be computed now that the mass of the Higgs boson is known.  And yet, this precise theory has no conceptual foundation.  Current measurements are consistent with the predictions of the SM
at their current level of accuracy, but deviations from the SM predictions at the 30\%, 10\%, or 3\% level are all possible in different highly plausible models.   The nature of the Higgs boson is a central part of the mystery of TeV physics. New physics responsible for Higgs condensation must couple to the Higgs boson and affect its properties at some level.

Full details of the future program on the Higgs boson, and more precise statements of the uncertainty estimates given below, can be found in the Higgs Boson working group report \cite{Higgsworking}.

\subsection{Higgs boson couplings} \label{Hcoupl}

The most direct question to ask about the new boson is whether it is in fact the sole source of mass for all quarks, leptons, and gauge bosons. For this to be true, the couplings of the Higgs boson to the various species of SM particles must follow a definite pattern.   The couplings of the particle to fermions and vector bosons must be, from Eq.~\ref{ff},
\beq   
 g_{Hf\bar f,SM} =   {m_f\over v} \ , \qquad     g_{HVV,SM} =   2 {m_V^2\over v} \ , \eeq{Higgsls} 
%\beq    \lambda_{hf\bar f,SM} =    {1\over \sqrt{2}}{m_f\over v} \ , \qquad     \lambda_{hVV,SM} =     {1\over 2} {m_V^2\over v^2} \ , \eeq{Higgsls} 
where $v$ is the value of the SM Higgs condensate, equal to 246~GeV. (More properly, this is the leading-order prediction for a coupling defined to all orders with all three particles on shell.) These couplings have a simple pattern that should be tested for as many SM species as possible.  In the following discussion, we define the scale factors
\beq 
 \kappa_A =   g_{HA\bar A}/(SM)  \  , 
\eeq{kappadef} 
%\beq   \kappa_A =   \lambda_{hA\bar A}/(SM)  \  , \eeq{kappadef} 
where $(SM)$ denotes the SM prediction.

The Higgs boson also couples to pairs of vector bosons $gg$, $\gamma\gamma$, and $\gamma Z$ through loop diagrams. In the SM, these couplings are dominated by contributions from $W$ boson and top quark loops.  In more general theories, these couplings can also receive contributions from radiative corrections with new particles in loops.  We will denote ratios of the on-shell couplings to the SM predictions by $\kappa_g$, $\kappa_\gamma$, and $\kappa_{\gamma Z}$.

Corrections to the predictions \leqn{Higgsls} can appear at many levels.  If there are multiple Higgs fields that mix into the observed boson, the $\kappa_A$ will contain cosines of the mixing angles.  These can be as large as the data permit.   Radiative corrections due to loop effects of new particles are expected to be below the 10\% level.

Corrections to the Higgs couplings are also affected by the decoupling theorem \cite{haber}: If all new particles have masses greater than $M$, we can integrate out these particles.  This leaves the SM, in which the properties of the Higgs boson are predicted precisely in terms of its mass.   The corrections to the SM values are generated by effective higher-dimension operators added to the SM Lagrangian. These corrections will then be at most of the order of $m_h^2/M^2$.   The decoupling theorem implies an apparently paradoxical but nevertheless important conclusion:  In a model in which the Higgs sector is very complex but all new particles in it are heavier than 500~GeV, corrections to the Higgs boson properties are at most at the 5-10\% level. We are likely to be in this situation, in which  the picture of the Higgs boson may be very different from that in the SM but, since the other particles in the sector are heavy, it is difficult to conclude this except by precision measurement.

Typical sizes of Higgs boson coupling modifications are shown in Table~\ref{tab:Htheorytable}.  More details of these estimates are given in \cite{Higgsworking}.

%%%%%%%%%%%%%%%%%%%%%%%%%%%%%%%%%%%%%
\begin{table}[htb!]
\begin{center}
\begin{tabular}{cccc}
\hline\hline
Model  &$\kappa_V$ & $\kappa_b$& $\kappa_\gamma$\\
  \hline
Singlet Mixing   &     $\sim 6$\%  & $\sim 6$\%  &  $\sim 6\%$ \\
2HDM &   $\sim 1\%$   & $\sim 10\%$  &  $\sim 1\%$ \\
Decoupling MSSM    &     $\sim -0.0013\%$ 
& $\sim 1.6\%$ & $< 1.5\%$\\
Composite  & $\sim -3\%$ & $\sim -(3-9)\%$ & $\sim -9\%$\\
Top Partner  &  $\sim -2\%$ & $\sim -2\%$ & $\sim +1\%$\\
 \hline\hline
\end{tabular}
\end{center}
\caption{Generic size of Higgs coupling modifications
from the Standard Model values in classes of new physics models:  mixing
of the Higgs boson with a singlet boson, the two-Higgs doublet model, 
the Minimal Supersymmetric Standard Model, models with a composite
Higgs boson, and models with a heavy vectorlike top quark partner. For these
estimates,
all new particles are taken to have  $M\sim 1$~TeV and mixing angles are 
constrained to satisfy precision
electroweak fits. }
\label{tab:Htheorytable}
\end{table}
%%%%%%%%%%%%%%%%%%%%%%%%%%%%%%%%%%%

Tests of the values of the Higgs couplings relative to the SM must take account of the theoretical uncertainty in the comparison to the SM predictions.  A potentially observable quantity is the partial decay width $\Gamma(h\to A\bar A)$, related to $\kappa_A$ by 
\beq      \kappa_A^2 = \Gamma(h\to A\bar A)/(SM)  \ . 
\eeq{kappaGamma} 
Currently, the SM predictions for the values of some Higgs partial widths have large uncertainties.   The uncertainty in the partial width $\Gamma(h\to b\bar b)$, which accounts for more than half of the SM Higgs total width, is quoted as 6\%~\cite{LHCHiggsref}.   A concerted program is required to bring the uncertainties in the  SM predictions below 1\%. This  requires complete evaluation of the 2-loop electroweak corrections to the partial widths.  It also requires improvement of the uncertainty in the crucial input parameters $\alpha_s$, $m_b$, and $m_c$. Lattice gauge theory promises to reduce the errors on all three quantities to the required levels \cite{LGTforHiggs}.  Further methods for improvement in our knowledge of $\alpha_s$ are discussed in Section \ref{sec:qcd}.

 \begin{figure}[htb]
\begin{center}
\includegraphics[width=0.45\textwidth]{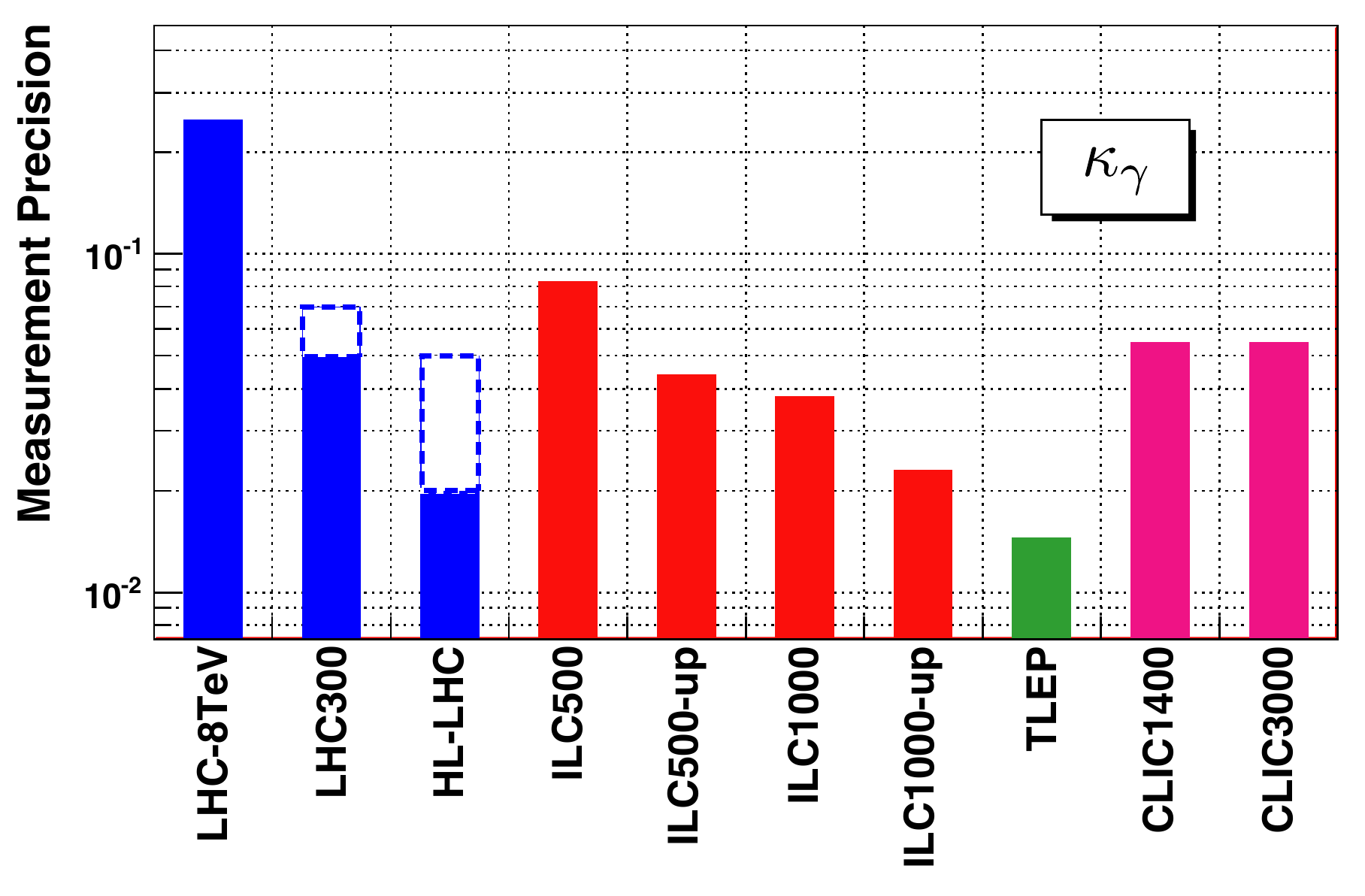}  \quad 
\includegraphics[width=0.45\textwidth]{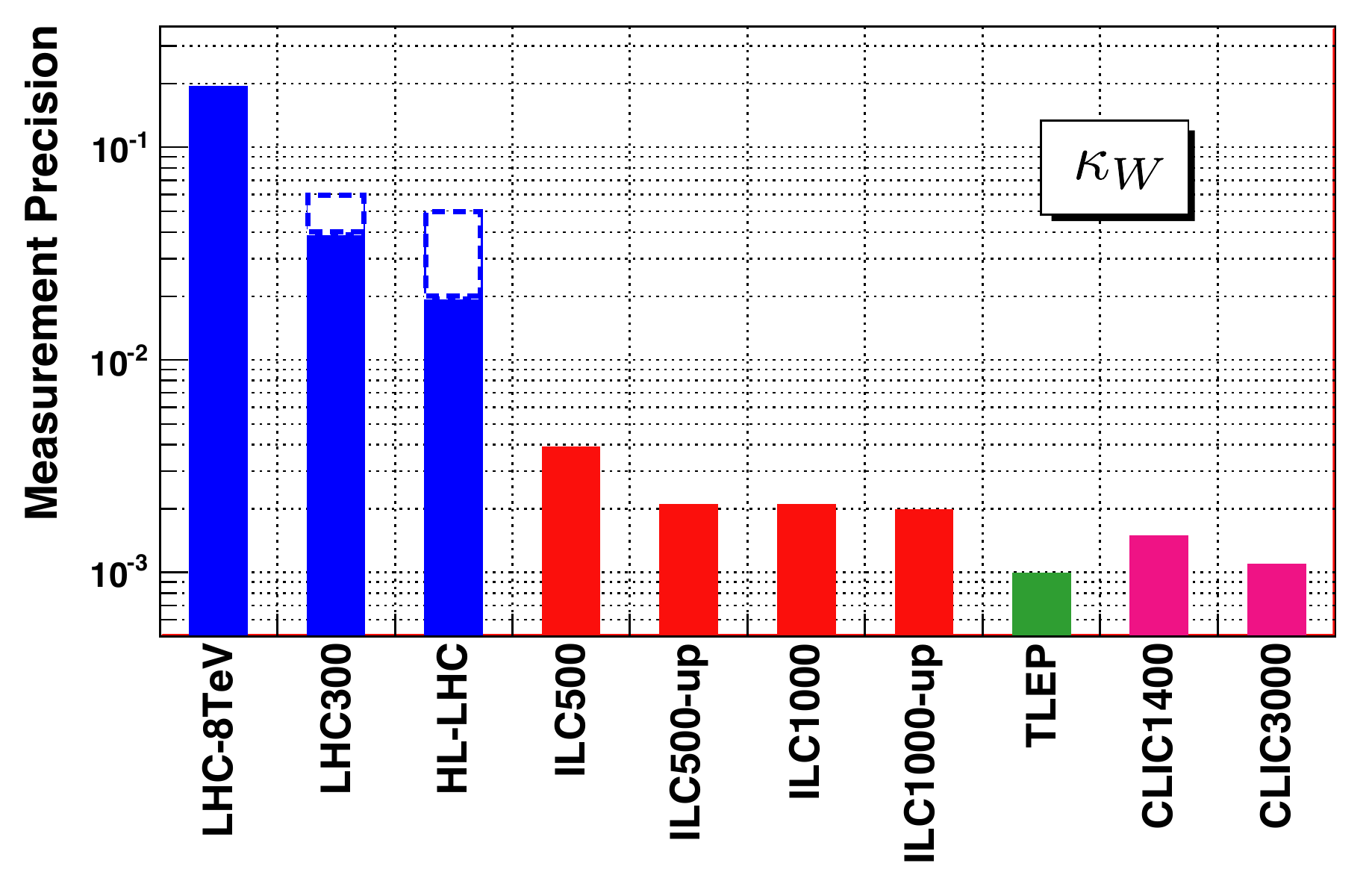} \\ 
\includegraphics[width=0.45\textwidth]{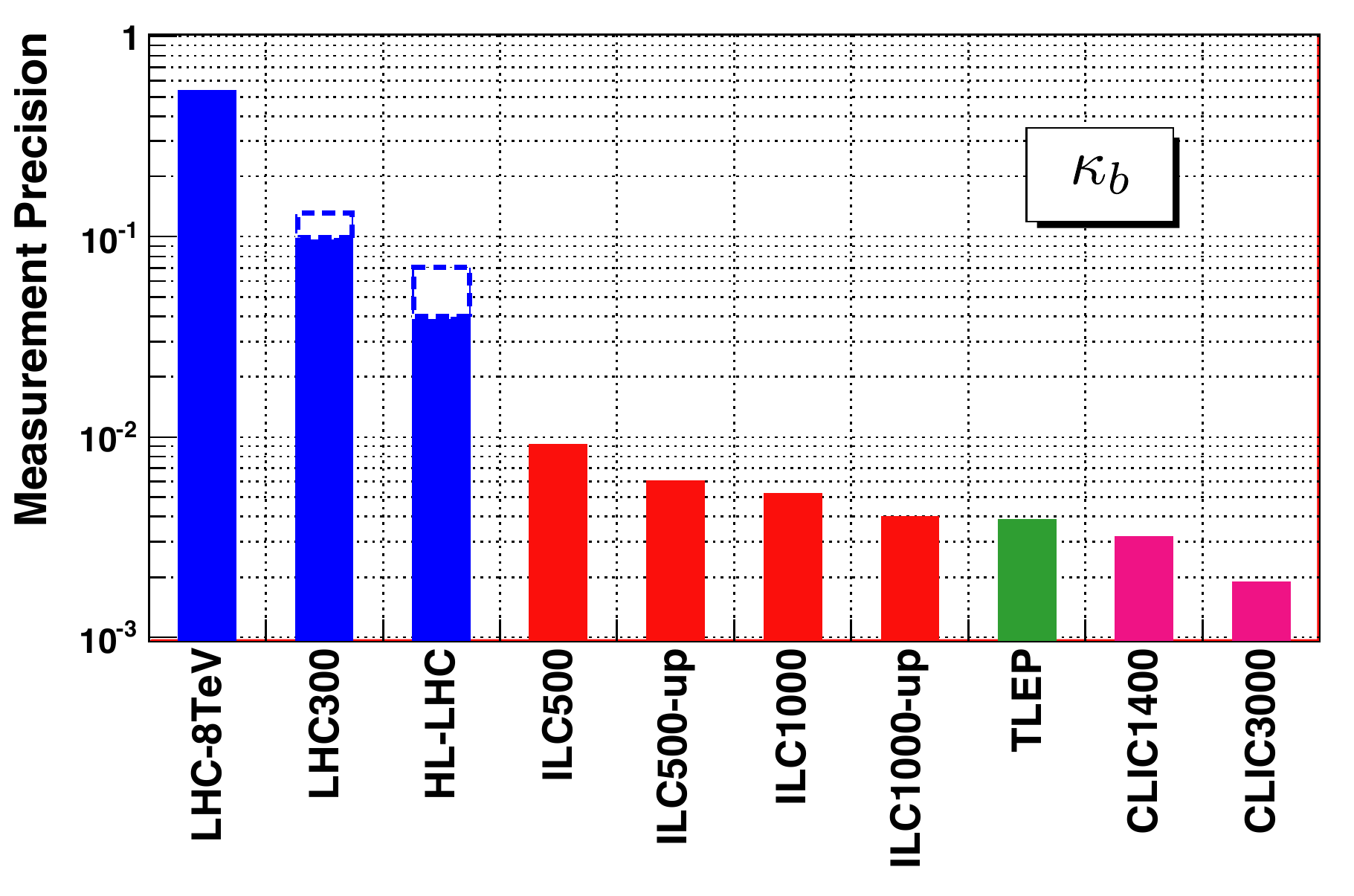}  \quad 
\includegraphics[width=0.45\textwidth]{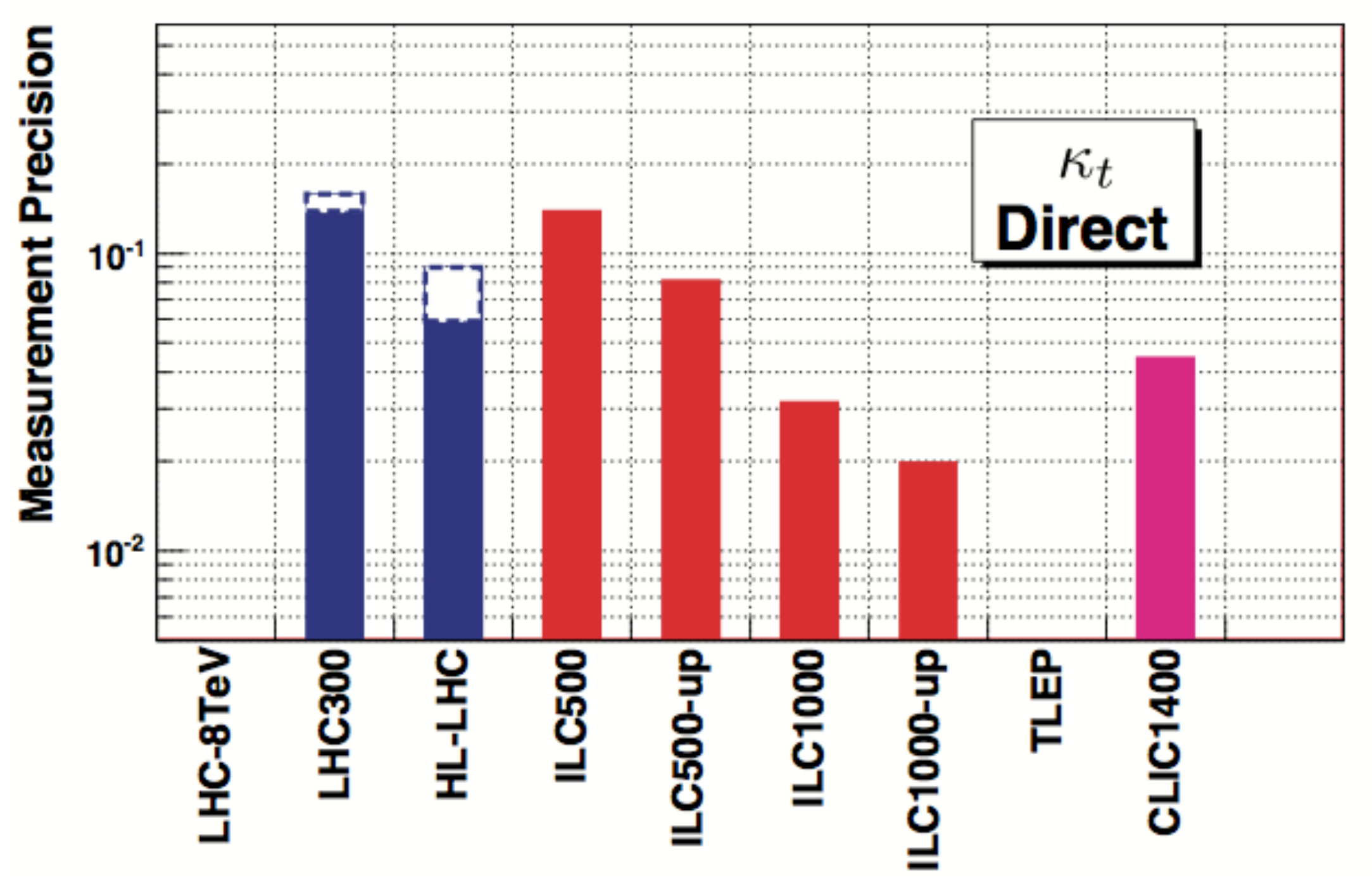} 
\end{center}
\caption{Expected measurement precision on Higgs boson couplings at different 
colliders.  The couplings considered are: (top row) $\kappa_\gamma$, $\kappa_W$, (bottom row)
$\kappa_b$, $\kappa_t$.  Blue bars correspond to stages of LHC, with the white band showing a range between
pessimistic and optimistic scenarios.  Red bars correspond to stages of ILC, including the 
projections for TDR luminosity samples and proposed  luminosity upgrades.  Green and purple
bars corresponds to TLEP and CLIC.  There is no TLEP entry in the $\kappa_t$ plot.} \label{fig:kappas}
\end{figure}

There are only a few cases in which the partial widths $\Gamma(h\to A\bar A)$ can be measured directly.  More often, the Higgs decay partial widths are measured from the rates of reactions that involve the Higgs boson in an intermediate state.  An example is the rate of $\gamma\gamma$ production through $gg$ fusion at the LHC.  The rate of this process is proportional to the Higgs boson  production cross sections times the 
branching ratio of the Higgs boson to $\gamma\gamma$,
\beq   
 \sigma(gg\to h) \cdot BR(h\to \gamma\gamma) \sim {\Gamma(h\to gg)      \Gamma(h\to \gamma\gamma)\over \Gamma_T(h) } \ , 
\eeq{sigmaBR} 
where $\Gamma_T(h)$ is the total Higgs boson width.   In terms of the $\kappa_A$ quantities, the measured rates are proportional to 
\beq    
  \sigma(A\bar A \to h) BR(h\to B\bar B) /(SM) =   {\kappa^2_A        \kappa^2_B\over \sum_C \kappa^2_C BR_{SM}(h\to C\bar C)} \  . 
\eeq{sigBRkappa} 
 The SM prediction for the total width of the Higgs boson is 4~MeV, a value too small to be measured directly except at a muon collider where the Higgs boson can be produced as a resonance.  At all other cases of hadron and lepton colliders, the total width must be determined by a fit to the collection of measured rates.  Such fits entail some model-dependence to control the size of modes of Higgs decay that are not directly observed.

The report \cite{Higgsworking} compares the abilities of experiments at  the LHC and at a variety of lepton colliders to extract the values of the Higgs boson couplings.  At the LHC, the total number of Higgs bosons produced is very high, over 170 million per experiment for integrated luminosity of 3000 fb$^{-1}$.   However, Higgs boson production at the LHC is accompanied by very high backgrounds. The discrimination of signal 
from background brings in substantial systematic uncertainties.  The extraction of couplings from cross sections is complicated by significant QCD uncertainties in the calculation of cross sections, currently about 12\% for gluon fusion and 3\% for vector boson fusion.

At electron colliders, the Higgs boson is produced in the relatively background-free processes $\ell^+\ell^- \to Zh$ and $\ell^+\ell^-\to  \nu\bar \nu h$ (vector boson fusion).  The systematic errors in the extraction of Higgs couplings are small. The main uncertainties come from limited statistics.  The $Zh$ reaction offers tagged Higgs bosons, giving the possibility of observing decay modes not accessible at the LHC (such as decay to $c\bar c$), and invisible and exotic modes of Higgs decay.   The total  cross sections for the two $\ee$  reactions  are directly proportional to $\Gamma(h\to ZZ^*)$ and $\Gamma(h\to WW^*)$, respectively, without dependence on $\Gamma_T(h)$.  This allows lepton collider measurements to determine $\Gamma_T(h)$ and all 
individual partial decay widths by fitting of Higgs boson rates  without any model assumptions.

Figure~\ref{fig:kappas}  compares the projected uncertainties
in the measurement of Higgs boson couplings for a variety of $pp$ and lepton collider programs.
 The first three figures show
the uncertainties in the couplings to  $\gamma\gamma$, $WW$,  and $b\bar b$ from a 6-parameter fit 
appropriate to the analysis of LHC results, described in \cite{Higgsworking}.  
Although the model-independent results from
lepton colliders do not require such a fit, the same fit is used for those cases
to facilitate comparisons to the hadron colliders.   
 The fourth figure shows the projected error on the 
Higgs coupling to $t\bar t$ from experiments directly sensitive to this quantity.  The
 facilities considered are the LHC at its current stage, the LHC after 300~\ifb\ and after 3000~\ifb, 
 the ILC up to 500 GeV and up to 1000 GeV, the TLEP circular $\ee$ collider, and the CLIC
 linear collider operating at 1400 and 3000 GeV.  For LHC, the upper and lower estimates
 reflect a pessimistic scenario, in which systematic errors not evaluated from data
do not improve, and an 
optimistic scenario, in which theory errors are halved and all experimental systematic errors
 decrease as the square root of the integrated luminosity,   For the ILC stages, the first error
 bar corresponds to the baseline event samples considered in the ILC TDR, while the second 
includes, more optimistically, a set of luminosity upgrades described in the TDR.   Full 
details, and tables of the numerical results of the fits, can be found in \cite{Higgsworking}. 
 The figures show that the LHC, especially in its high luminosity phase, will measure
 Higgs couplings with an impressively high precision of several percent. 
However, the discovery of  perturbations 
of the Higgs boson couplings at the level shown in Table~\ref{tab:Htheorytable}, at 
3--5$\sigma$ significance, will need even higher precision.  This will require 
both the much lower level of systematic errors 
available at a lepton collider and very large event samples  to reduce the statistical errors.

\subsection{Higgs boson self-coupling} \label{Hscoupl}

A particularly important coupling of the Higgs boson is the Higgs self-coupling,  $\lambda$ in \leqn{higgspot}, which determines the shape of the Higgs potential.  In the SM, after Higgs condensation, there is a triple Higgs boson  coupling proportional to $\sqrt{\lambda}$, given alternatively by 
\beq         
    \lambda_{hhh} =   {3 m_h^2\over v} \ . 
\eeq{tripleHiggs} 
This coupling can be extracted from the rate for double Higgs production, for example $pp \to hh + X$ or $\ee \to \nu\bar\nu hh$.

Theoretical models with extended Higgs sectors or composite Higgs bosons can predict that the triple Higgs coupling will deviate from the SM expectation by 20\%.   This is predicted to be a larger effect
 than those expected in the Higgs couplings to fermions and vector bosons. However,
 the measurement is also much more difficult.   The cross sections at lepton colliders are at the fb level.  
At the LHC, the cross sections are larger, but it may be necessary to detect the Higgs boson in a rare
decay mode such as $\gamma\gamma$ in order to reduce the background. 
 Studies using only a few decay modes indicate a precision of 50\% in 
the Higgs self-coupling measurement per
detector. By combining the results of the detectors, the HL-LHC program would likely provide 
the  first evidence of the Higgs self-coupling.
 The projected uncertainties on the Higgs self-coupling are  13\% in a long-term program at the ILC at 1~TeV or 10\% for CLIC at 3~TeV.   The double Higgs production cross section increases rapidly with energy.  Measurements at a 100~TeV $pp$ collider are estimated to reach an uncertainty of 8\%.

\subsection{Higgs boson spin and CP} \label{Hspin}

A crucial test of the identification of the 125~GeV resonance with the Higgs boson is the measurement of its spin and parity.  This issue is almost settled with the current data from the LHC.  The fact that the resonance decays to $\gamma\gamma$ implies that it has integer spin and cannot have spin 1.   The distribution of the four leptons in $h\to ZZ^*$ decays already strongly favors the $0^+$ over the $0^-$ spin-parity hypothesis and excludes  the simplest forms of spin 2 coupling~\cite{CMSHiggsspin,ATLASHiggsspin}.  
This issue should be decided with the next LHC data set.

However, there is a more subtle issue associated with the Higgs boson CP.  If there are multiple Higgs bosons and CP violation in the Higgs sector, the Higgs boson at 125~GeV can contain an admixture of CP scalar and pseudoscalar states.  
CP violation in the Higgs sector has major implications. Most importantly, it can provide the new source of CP violation outside the SM that allows the matter-antimatter asymmetry of the universe to be generated at the electroweak phase transition.

CP violation in the Higgs sector can be reflected both in production and decay of the 
Higgs boson.   The most accurate tests are available in the study of the 4-lepton 
final state in $h\to ZZ^*$. The scalar Higgs couplings to massive vector bosons
appear at tree level, while pseudoscalar couplings are expected to
appear only in loop corrections.  CP-violating terms in this vertex are therefore
masked by the large CP-conserving tree-level amplitude.
However, measurements in vector boson fusion and in 
associated production are potentially accurate enough to overcome the loop suppression.
 Lepton colliders can search for CP violation 
in the decay $h\to \tau^+\tau^-$ and in the production process $\ell^+\ell^- \to t\bar t h$.
 The first process can reach 1\% precision in the measurement of a cross-section fraction corresponding 
to the CP-odd decay amplitude~\cite{htotautau}.

 Photon-photon colliders, which produce the Higgs boson as a 
resonance, can use initial-state polarization to search for CP-violating terms in the
 Higgs boson coupling to $\gamma\gamma$, which has no tree-level contributions.  
Similarly, a muon collider can probe for CP-violating contributions to the Higgs 
boson coupling to $\mu^+\mu^-$ if the accelerator provides transverse beam  polarization.

\subsection{Higgs boson mass and width} \label{Hmass}

The Higgs boson mass is currently known from the LHC experiments to better than 600~MeV.   This accuracy is already sufficient for the uncertainty in the Higgs mass  not to be significant in precision electroweak tests.  The most important influence of a highly accurate Higgs mass within the SM comes in the evaluation of the predictions for the Higgs couplings to $WW$ and $ZZ$, for which one boson must be off the mass shell.   A 100~MeV error in the Higgs mass corresponds to a 0.5\% uncertainty in $\kappa_W$. We expect that the  error in the Higgs mass can be decreased to 100~MeV  and to 50~MeV, respectively, for the LHC programs with 300 fb$^{-1}$ and 3000~fb$^{-1}$ by using the $\gamma\gamma$, $ZZ^*$, and $\mu^+\mu^-$ modes, in which the Higgs boson can be fully reconstructed.    A lepton collider studying the Higgs boson in the $Zh$ production mode would push this uncertainty down further, to about 35~MeV  for linear colliders and 7~MeV  for a very high luminosity program at a circular collider.

Predictions of the Higgs mass in models of new physics might provide further motivation for measuring the Higgs mass accurately.  An example of such a model is the Minimal Supersymmetric Standard Model (MSSM). To evaluate the prediction to an accuracy of 100~MeV, however, the masses of the top squarks must be known, and the top quark mass must be known to 100~MeV.

We have noted already that lepton colliders offer the possibility of a model-independent determination of the Higgs boson total width. Because the couplings of the Higgs boson to $ZZ$ and $WW$ appear in both the expressions for measurable total cross sections and those for
 branching ratios, these couplings can be eliminated to evaluate the total width through the relations 
\beq      \Gamma_T(h) \sim  {   \sigma(\ell^+\ell^- \to Zh)/BR(h\to ZZ^*)}      \sim { \sigma(\ell^+\ell^- \to \nu\bar\nu h , h\to b\bar      b)/BR(h\to WW^*)BR(h\to b\bar b)} \eeq{Gammaeval} 
This gives the Higgs boson width to 3\% for a long-term program at the ILC and to 0.6\% for a high luminosity program at a circular collider with multiple detectors.   These uncertainties are reflected in the coupling uncertainties quoted in Section \ref{Hcoupl}.

A muon collider would have the capability of observing the Higgs boson as a narrow resonance.   For the projected beam energy resolution of $4\times 10^{-5}$, the mass of the Higgs boson would be measured to 0.06~MeV and the width would be measured directly in the $s$-channel to a precision of 4\%~\cite{MuonHiggs}

\subsection{Searches for additional Higgs bosons} \label{Haddl}

There are strong motivations for expecting the existence of additional 
Higgs particles.  These motivations begin with the overall mysteries of  
the physics of Higgs condensation and the question of whether the Higgs 
boson is the only particle of the SM whose quantum numbers do not come
 in multiples.  Beyond this, virtually all models of new physics to explain the 
Higgs potential contain additional Higgs doublet fields. These fields are 
required in supersymmetric models in order for Higgs fields give mass to 
both the up-type and the down-type quarks.  In models with new space
 dimensions, additional Higgs fields arise as the Kaluza-Klein excitations 
of the fundamental Higgs 
doublet.    Each additional Higgs doublet gives rise to four new particles, 
CP-even and CP-odd neutral scalars $H$ and $A$, and a charged pair $H^\pm$.

  Often, extended Higgs bosons have enhanced couplings to heavy flavors, 
either to $b$ quarks and $\tau$ leptons or to $t$ quarks, depending on whether the extended Higgs 
parameter $\tan \beta$ is greater than or less than 1.   This emphasizes
 searches in which the extended Higgs bosons are produced in 
$b\bar b$ annihilation.

Currently, the LHC experiments based on heavy flavor  signatures exclude 
additional Higgs bosons for masses as high as 1 TeV in restricted ranges of $\tan\beta$. 
The region of large $\tan\beta$ is surveyed by reactions such as $b\bar b \to  
 H, A \to \tau^+\tau^-$, while the region of  low $\tan\beta$ is surveyed by reactions 
such as $gg\to H, A \to t\bar t$, $gg\to A \to Zh$.   A gap remains for intermediate 
values, roughly $2 < \tan\beta < 20$, which is closed only if the  extended Higgs bosons 
have masses below 200~GeV. Future runs of the LHC, up to 3000~fb$^{-1}$, are expected to 
close this window up to masses of about 500 GeV.

\begin{figure}
\centering
\includegraphics[width=0.48\linewidth]{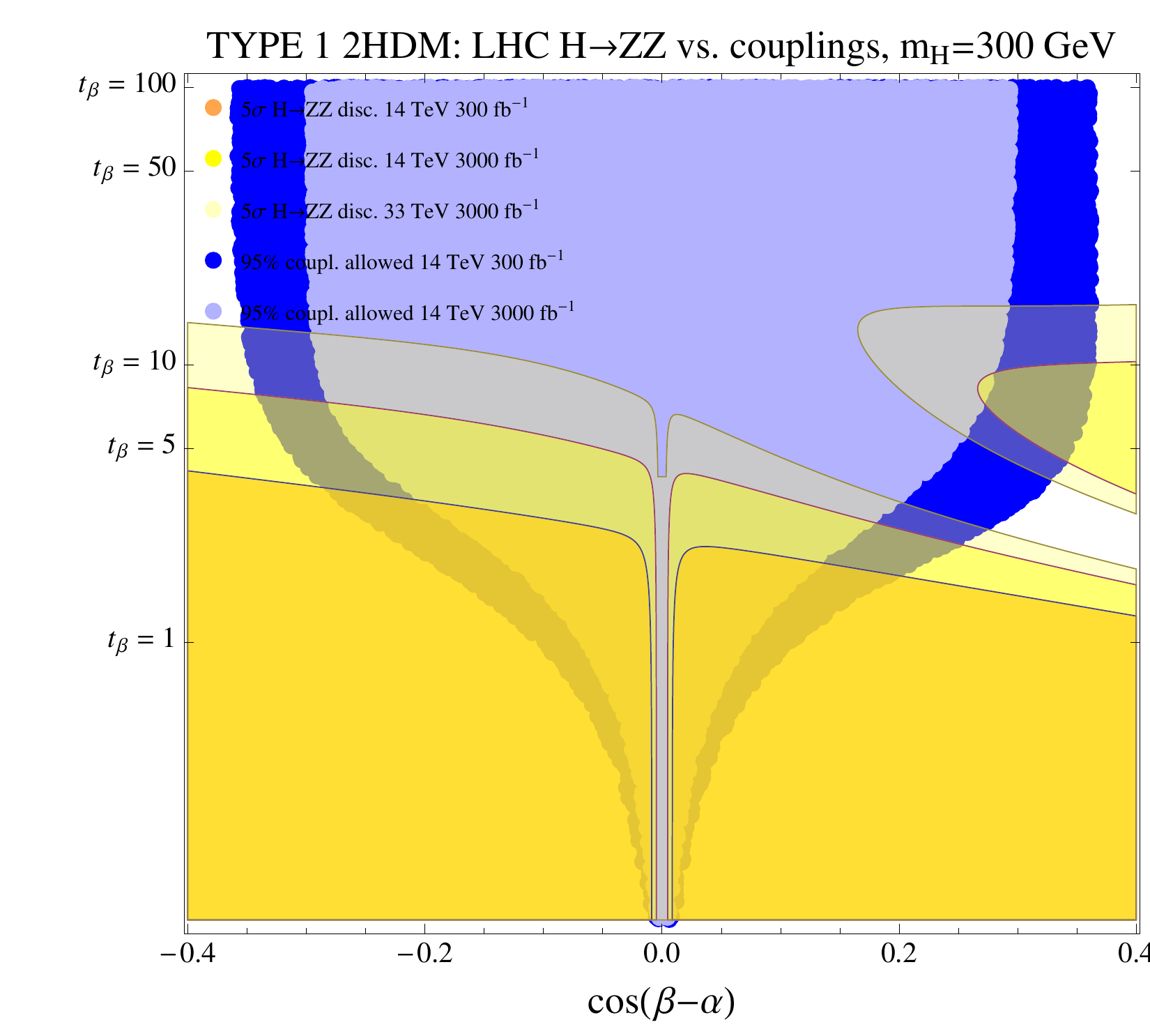}
\includegraphics[width=0.48\linewidth]{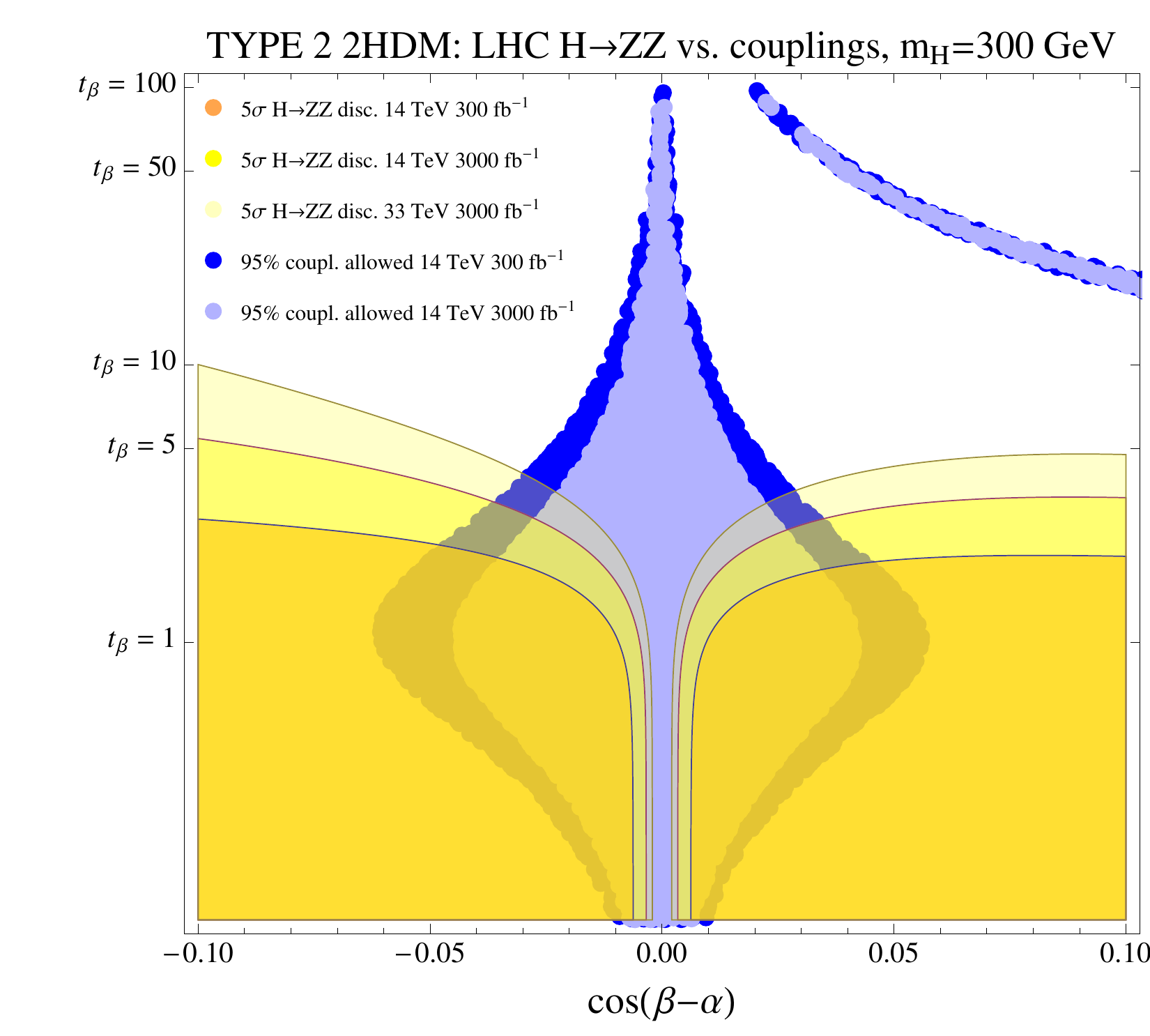}
\caption{$5\sigma$ discovery reach for  a 300~GeV $H$
decaying via $H \rightarrow ZZ \rightarrow 4\ell$, for the Type I
and Type II two-Higgs-doublet models, shown in the left and right
panels, respectively.   The coordinates are  the combinations of 
Higgs boson mixing angles $\cos(\beta-\alpha)$
and $\tan\beta$.  The yellow regions show the 5 $\sigma$ reach in 
direct searches at the LHC.  The successive larger regions correspond to
the LHC at 300~\ifb\ and 3000~\ifb\ and a 33 TeV $pp$ collider with 
3000~\ifb.   The blue regions show the regions allowed at
95\% CL by precision Higgs coupling measurements;  the 
darker and lighter regions correspond to the LHC with 300~\ifb\ and
3000~\ifb~\cite{Barger:2013ofa}.}
\label{fig:HZZreach}
\end{figure}

Additional Higgs particles are typically predicted to  have smaller  couplings to $WW$ and $ZZ$ 
than the lightest Higgs boson. These couplings are  proportional to the combination of  mixing angles $\cos(\beta-\alpha)$, which
 are also constrained by measurements of the vector boson couplings to the known Higgs boson.
The complementarity of these search strategies is illustrated with a search
for additional Higgs bosons in a model with two Higgs doublets \cite{Barger:2013ofa,ChenHA,2HDMwp}.
Fig.~\ref{fig:HZZreach} shows the discovery reach for a 300~GeV Higgs boson $H$
decaying via $H \to ZZ \to 4\ell$ compared with the allowed region from Higgs
coupling measurements at various experimental facilities.
The reach for a pseudoscalar $A$ decaying via $A \to Zh \to \ell \ell b \bar b, \ell \ell \tau \tau$ is
comparable~\cite{2HDMwp}.
The important parameters for this search are the Higgs mixing angle $\alpha$ and
the ratio of the Higgs vacuum expectation values $\tan\beta$.  The efficacy of the search
also depends on which Higgs
bosons couple to which quarks and leptons. (This distinguishes the Type I and Type II
models in the figure.)
The line $\cos(\beta - \alpha) = 0$ is the limit where the additional
Higgs particles decouple from gauge bosons and the couplings of the 125 GeV 
Higgs boson are very close to the predictions of the SM. 
Even for Type II models (which include supersymmetry), where the Higgs coupling
measurements are already very constraining, the direct Higgs search probes
significant additional parameter space.
There is roughly a factor of 2 increase in the reach for large $\tan\beta$ in
each step in going from LHC at 300~\ifb\ to HL-LHC to a 33 TeV collider with
3000~fb${}^{-1}$.

Lepton collider experiments can search for extended Higgs boson states through 
the reaction $\ell^+\ell^- \to HA$ up to the kinematic limit, independently of the 
value of $\tan\beta$.   The cross section depends only on the electroweak quantum 
numbers of the extended Higgs particles. This covers the parameter space up to 
500~GeV for ILC at 1~TeV and up to 1500~GeV for CLIC running at 3~TeV.  Photon and 
muon colliders have the opportunity to discover additional Higgs bosons as 
resonances up to the full center of mass energy of the machine.

% %
%\line(1,0){\textwidth}

% \subsection{The Message} \label{Higgsm} % %

% The conclusions of the Higgs Boson working group can be summarized as follows: 
% \begin{enumerate} 
% \item Direct measurement of the Higgs boson is the key to   understanding electroweak symmetry breaking.   The fact that the Higgs boson  appears as a light, apparently fundamental, scalar particle needs  explanation. A research program focused on the Higgs couplings to fermions and vector bosons and achieving a precision of a few percent or less is required to address these questions. 
% \item  Full exploitation of the LHC is the path to few percent   precision in the Higgs coupling and to a 50 MeV precision in the determination of the Higgs mass. 
% \item Full exploitation of a precision electron collider is the path to a model-independent measurement of the Higgs boson width and a sub-percent measurement of the Higgs couplings. Such precision is necessary to 
% probe for new physics beyond the reach of LHC direct searches.
%  \end{enumerate}
%  Experiments on Higgs bosons give information on the Particle Physics  Questions \# 1, 2, 4, 5, 8, 9, 10 listed in the Snowmass Summary~\cite{Summary}.

% % %
% \line(1,0){3.2in}
% \vspace{-12pt}
 \section{Electroweak Interactions} \label{sec:EW} %
% \vspace{-12pt}\line(1,0){3.2in}
% %
% \vspace{12pt}
% %

All particle species with couplings to the electroweak interactions eventually influence the properties of the weak interaction bosons $W$ and $Z$. Very precise measurements of the properties of these bosons then have the potential to reveal new, undiscovered particles. The precision electroweak experiments of the 1990's established the $SU(2)\times U(1)$ theory of electroweak interactions at the sub-percent level of accuracy.  But, also, they indicated the presence of a heavy top quark and a light Higgs boson and estimated the masses at which these particles were eventually discovered.  They disfavored a fourth generation of quarks and leptons, now excluded by direct search at the LHC.

Increased precision in the properties of the weak interaction bosons could well turn up the first evidence of the TeV spectrum of particles discussed in Section~\ref{sec:EF-TeV}.  Some observed deviations from the
predictions of the SM have remained intriguing as the measurements have improved.  For example, the measured  value of $M_W$ has persistently remained  1--2 $\sigma$ higher than the SM expectation.  Experiments over the next decade will explore whether these deviations  could become significant 
effects requiring radiative corrections due to new particles.

In a description of possible new interactions in terms of effective operators, the electroweak precision observables probe only the first few terms. Experiments at higher energy probe additional operators by observing and constraining the nonlinear interactions of the $W$ and $Z$ bosons.  These operators can receive corrections from loop diagrams involving new TeV mass particles but, more strikingly, they can receive leading-order corrections if there is new strong dynamics or resonances in the Higgs sector.

We will review these topics in this section.  Full details of the program, and more precise statements of the projected uncertainties described below, can be found in the Electroweak Interactions working group report \cite{EWworking}.

\subsection{Precision observables $M_W$ and $\ssteff$} \label{Wmass}

The current uncertainty in  the $W$ boson mass is 15~MeV, corresponding to a relative precision of $2\times 10^{-4}$.  Currently, the most accurate determinations of $M_W$ come from the hadron collider experiments CDF and D$\dzero$.
Precision measurement of $M_W$ at hadron colliders is very challenging, but certain features of 
$W$ boson production make it feasible to reach high accuracies.  The directly measured transverse mass distribution is very sensitive to $M_W$, having a relatively sharp endpoint at the $W$ boson mass. Likewise the transverse momentum 
distributions of the leptons are also sensitive to the $W$ 
boson mass with different, but manageable, systematic uncertainties. The $W$ boson production 
cross section is large, and the event samples generally have  very small contamination by background. 
 Currently, the largest source of systematic error is the dependence of the acceptance on the rapidity of the produced $W$ boson, requiring a correction that depends on quark and antiquark parton distribution functions (PDFs). Experimental uncertainties are at the same level as those due to PDFs and are expected to continue to decrease accordingly.

We see good prospects for improving the measurement of $M_W$ 
at the LHC. The statistical component of the error will be negligible already
 with the current LHC data set.  The error from PDFs doubles in going from the Tevatron to the LHC because proton-proton collisions involve
 no valence antiquarks.  However, we anticipate that this error will be decreased using new data on the vector boson rapidity and charge asymmetries.  The issue of PDF improvement is discussed further in Section~\ref{QCDPDFs}. The huge statistical precision available at the LHC 
will allow for control of calorimetric and tracking systematic uncertainties.
 In Table~\ref{tab:mwerror} we see that the PDF error in $M_W$ is expected to be 
brought down to $\pm$ 5~MeV~ with 300~\ifb\ and to $\pm$ 3~MeV~ with 3000~\ifb. For this last estimate, the uncertainties in the PDFs must be
pushed to a factor of 7 smaller than those today. 
 At each stage, the experimental systematics are expected to keep pace with the improvements in the PDFs, leading to a final uncertainty in $M_W$ of $\pm$ 5~MeV.

%%%%%%%%%%%%%%%%%%%%%%%%%%%%%%%%%%%%%%%%% 
\begin{table}[t]
\begin{center}
\begin{tabular}{l|rrr}  
$\Delta M_W$ [MeV] & \multicolumn{3}{c}{LHC}  \\ \hline 
$\sqrt{s}$ [TeV] & 8  & 14  & 14  \\
$\L [\fb]$ & 20   &300  & 3000  \\ \hline
PDF            & 10 & 5 & 3   \\
QED rad.       & 4 & 3  & 2   \\
$p_T(W)$ model &  2 & 1 & 1  \\
other systematics &  10 & 5 & 3    \\
$W$ statistics    &  1 & 0.2& 0   \\
Total             &  15 & 8 & 5   \\
\end{tabular}
\caption{Current and target uncertainties (in MeV) for individual 
components of the precision
 measurement of 
$M_W$ at the LHC.}
\label{tab:mwerror}
\end{center}
\end{table}

%%%%%%%%%%%%%%%%%%%%%%%%%%%%%%%%%%%%%%%%%%

Lepton colliders offer an opportunity to push the uncertainty in $M_W$ down even further.   The $W$ mass was measured at LEP to $\pm$ 36~MeV~ from  the kinematics of $W^+W^-$ production.   The uncertainty was dominated by statistical errors, with a substantial additional contribution  from the modeling of hadronization.  Both sources will benefit from the data set on $W^+W^-$, about 1000 times larger, that will be available  at next-generation $\ee$ colliders such as ILC and TLEP. We estimate an error below $\pm$ 4~MeV from this method, and a similar error from independent measurements on single $W$ boson production.

The ultimate $W$ mass measurement would come from a dedicated energy scan of the $W^+W^-$ threshold at 160~GeV.   Such a measurement could reach $\pm$ 2.5~MeV~ with the statistics available from the ILC and $\pm$ 1~MeV~ with the statistics available from TLEP.   At this level, systematic errors become dominant.  The program also requires a detailed precision theory of the $W^+W^-$ threshold, using methods now applied to the $t\bar t$ threshold.

The measurement of the value of the weak mixing angle $\ssteff$ associated with 
quark and lepton couplings to the $Z$ boson resonance offers an orthogonal probe of 
the electroweak interactions.  The current accuracy in $\ssteff$ is at the $16\times 10^{-5}$ 
level of precision and is dominated by measurements from LEP and SLC.  This level might be 
reached but probably will not be surpassed at the LHC.  Again, uncertainties in PDFs give
 the limiting systematic error.  Measurements from the polarization-dependence of the $Z$ 
cross section and from the $b$ quark forward-backward asymmetry are discrepant by about 
$3\sigma$, indicating an experimental question that should be resolved.

Future lepton colliders give an opportunity to improve the precision in $\ssteff$.   The ILC program includes a few months of running at the $Z$ resonance to produce a data set of $10^9$ $Z$'s, improving the statistics from LEP by a factor of 100 with highly polarized beams. The ILC detectors should also dramatically improve the capability for heavy flavor tagging.  This ``Giga-Z'' program should improve the uncertainty in $\ssteff$ by a factor of 10.  The program also would give new measurements of other $Z$ pole observables sensitive to new TeV mass particles, most importantly, the fraction $R_b$ of $Z$ decays to $b\bar b$.

TLEP envisions a multi-year program at higher luminosity to collect $10^{12}$ events on the $Z$ resonance. This potentially pushes the precision of electroweak measurements by another order of magnitude, though systematic contributions to the errors must still be understood.  Among other factors, the $Z$ mass must be measured more accurately than the current 2.5~MeV.  This is possible at TLEP if transverse polarization can be achieved in single-beam operation.  A direct measurement of $\ssteff$ is most effectively done using 
 longitudinal polarization in colliding beam mode. The feasibility of achieving this
 at TLEP needs to be understood.

Loop effects from TeV mass particles can produce shifts from the SM expectations
 at the $10^{-4}$ level in both $M_W$ and $\ssteff$, so the improved capabilities for precision electroweak measurements may give significant evidence for new particles.  Quantitative estimates of these shifts for a number of new physics models are given in \cite{EWworking}.

\subsection{Interactions of $W$ and $Z$ bosons} \label{threefourW}

The interactions of $W$ and $Z$ bosons can be studied through the measurement of vector boson pair production and multi-vector boson production.  This study has already begun at LEP and the Tevatron, where parameters of the triple gauge boson interactions were bounded within a few percent of their SM values.

Vector boson interactions are described in a unified way through the formalism of effective Lagrangians.  
In this formalism, we parametrize departures from the SM Lagrangian, in which the Yang-Mills vertices for $\gamma$, $W$, and $Z$ appear as terms of dimension 4,   by adding terms with operators of higher
dimension that are  invariant under the $SU(2)\times U(1)$ gauge symmetry.  A typical term involving an operator of dimension 6 is 
\beq    \delta \L = { c_W\over \Lambda^2} (D_\mu\Phi)^\dagger W_{\mu\nu} (D_\nu\Phi)  \ , \eeq{examplesix} 
where $\Phi$ is the Higgs doublet field and $W_{\mu\nu}$ is the $W$ boson field strength.  The overall size of the interaction is controlled by the parameter 
$\Lambda$, which has the dimensions of GeV and gives the energy scale of the new physics that is added
to the SM. The coefficient
$c_W$ is a number of order 1 giving the size of this interaction relative to those of other dimension 
6 operators.  The term \leqn{examplesix} and other similar, terms  contribute to the triple and quartic 
gauge  boson vertices.  Additional operators of dimension 8 can modify the quartic gauge boson  interactions independently of the triple-boson interactions. A typical term with a dimension 8 operator is
 \beq    \delta \L = { f_{T,0}\over \Lambda^4} \tr (W_{\mu\nu})^2 \tr    (W_{\lambda\sigma})^2 \ . 
\eeq{exampleeight} 

In a weak-coupling theory such as the SM, the coefficients $c_i$ and $f_j$ are induced by loop diagrams and should be highly suppressed by powers of $\alpha_w/4\pi\sim 10^{-3}$.  However, in theories with strong interactions in the Higgs sector, the $c_i$ and $f_j$ coefficients could be of order 1.  In this context, 
the $\Lambda$ parameters would be interpreted as the masses of Higgs sector resonances.  
For example, the operator \leqn{exampleeight} would be induced by a scalar resonance in the Higgs sector.

The current bounds on triple gauge boson couplings imply that the $\Lambda$ parameters associated with dimension 6 operators are higher than about 600~GeV.   High-statistics measurements of the triple gauge bosons by observation of $W^+W^-$ and $ZZ$ production in $\ee$ reactions at 500~GeV are expected to be sensitive to deviations from the SM that are 10 times smaller, pushing the sensitivity to $\Lambda$ almost to 2~TeV.

It is difficult for hadron colliders to have similar sensitivity to triple gauge couplings.  One source of this difficulty is that the LHC experiments study diboson reactions at higher energies, where additional terms from higher-dimension operators are important and so the extraction of the coefficients of dimension 6 operators is model-dependent. But there is a compensatory advantage.  Working at higher energy, the LHC will  study $W$ and $Z$ bosons at energies where Higgs sector strong interactions can dramatically alter the amplitudes for vector boson scattering and triboson production.

Some quantitative examples are presented in \cite{EWworking}. In examples studied there, the sensitivity to the coefficients $ f/\Lambda^4$ of dimension 8 operators implies that, assuming $f=1$, the LHC at
 300~\ifb\ would achieve exclusion of a $\Lambda$ value greater than 1.5 TeV. The HL-LHC, with 3000~\ifb, would roughly triple the significance of the effect, allowing 5 sigma discovery at 1.5 TeV or exclusion up to 1.8 TeV. If the origin of this new physics lies in strongly-coupled dynamics, the HL-LHC provides discovery-level sensitivity to virtual effects of new resonances with masses ($4\pi\Lambda$) up to 19 TeV. If interpreted as sensitivity to $f$, the HL-LHC is more sensitive by a factor of 2-3 compared to the LHC at 300~\ifb.

Increasing the energy allowed for the diboson system dramatically increases the physics reach. For discovery of anomalous couplings, a 33 TeV  $pp$ collider would reach $\Lambda$ values above 1.8 TeV, while a VLHC at 100 TeV would reach above 4.6 TeV.  These values extend well into the region in which new Higgs sector dynamics would be expected in models of this type.

% \subsection{The Message} \label{ewm} 
% The conclusions of the Electroweak Interactions working group can be summarized as follows: \begin{enumerate} 
% \item Precision measurements of the $W$ and $Z$ bosons has the potential to probe indirectly for new particles with TeV masses.  This precision program is within the capabilities of LHC, linear $\ee$ colliders, and TLEP. 
% \item  Measurement of vector boson interactions will  probe for new dynamics in the Higgs sector. In such theories, we expect correlated signals in triple and quartic gauge couplings.  The LHC and linear colliders will have sensitivity into the mass region above 1 TeV. 
% \end{enumerate} Experiments on electroweak interactions give information on the Particle Physics  Questions \# 1, 4, 8, 9 listed in the Snowmass Summary~\cite{Summary}.

% %
% % %
% \line(1,0){6in}
% \vspace{-12pt}
 \section{Quantum Chromodynamics and the Strong Interaction} \label{sec:qcd} %
% \vspace{-12pt}\line(1,0){6in}
% %
% \vspace{12pt}

%

Every probe of particle physics at high energies eventually requires detailed knowledge of the strong interactions.  Even in pure electroweak processes, the strong interactions affect the values of the 
electroweak coupling constants 
$\alpha$ and $\alpha_w$ through coupling constant renormalization.  
In the next decade, when most of our new knowledge about particle physics
 will come from hadron collider experiments, our understanding of the strong
 interactions will influence every aspect of the data.  Observables measured at the LHC 
are affected by  the structure of the
 proton, by radiation from initial- and final-state quarks and 
gluons,  by the transition from quarks to hadrons, and by 
 the detailed physics that produces multi-jet events.

The strong interactions are known to be described by the Yang-Mills theory
 Quantum Chromodynamics (QCD).   This theory has a coupling constant that is weak at  
short distances and strong at large distances. Our understanding of QCD 
is imperfect.  We have limited tools for the strongly coupled regime, and precision 
calculation in the weakly coupled regime is technically complex.  Nevertheless, 
our knowledge of QCD has taken enormous strides since the previous Snowmass 
workshop a decade ago.  In this section, we review the currents state of our tools
 for QCD and indicate the opportunities for further progress. More details on all 
of the topics discussed here can be found in the working group report \cite{QCDworking}.

In our discussion of precision quantum field theory calculation, we will describe one-loop radiative corrections as next-to-leading order (NLO), and higher corrections as NNLO, {\it etc.}

\subsection{Parton distribution functions} \label{QCDPDFs}

Our knowledge of the initial state in hadron-hadron collisions is encoded in the representation of the proton structure given by the parton distribution functions (PDFs).   The provision of PDF distributions with uncertainties is an innovation of the past decade \cite{Tung:2002vr}.  The  uncertainties quoted have been continually improved through the addition of new data sets, especially from the Tevatron and HERA experiments.

%%%%%%%%%%%%%%%%%%%%%%%%%%%%%%%%%%%%%%%%% 

\begin{figure}[htb] \begin{center} 
\includegraphics[width=0.48\hsize]{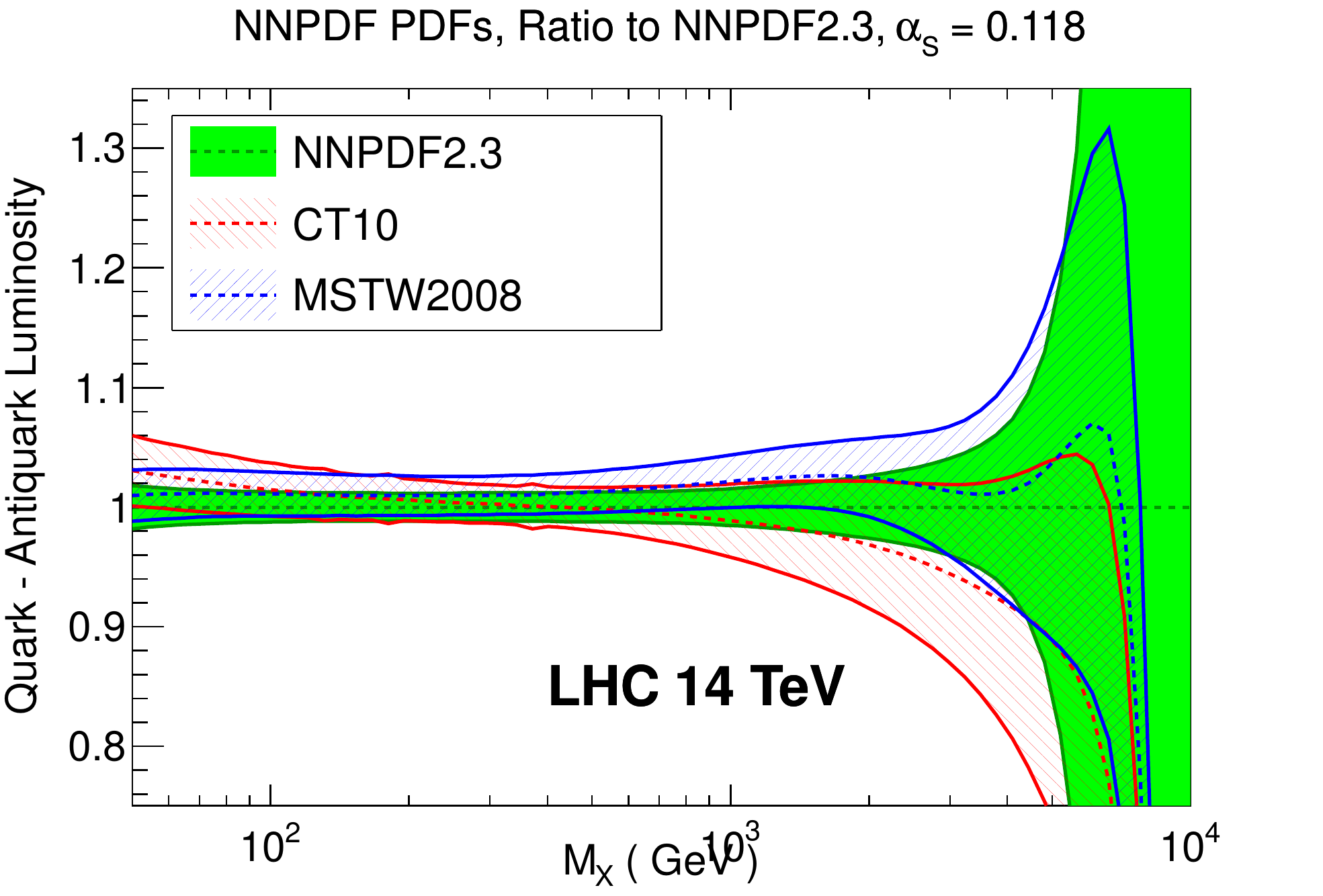}\ \ 
\includegraphics[width=0.48\hsize]{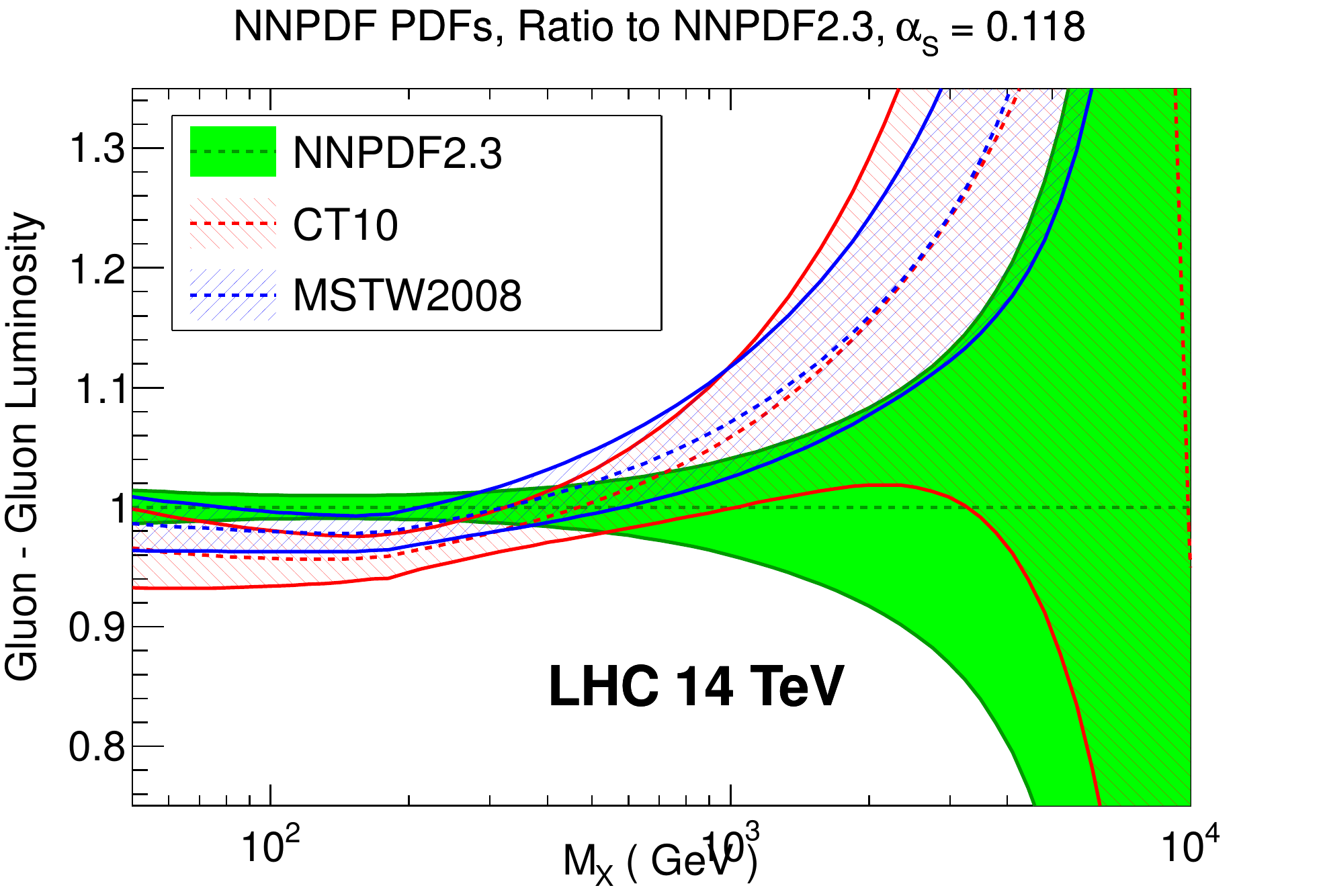}\ \ 
 \end{center}
 \caption{Comparison of the  partonic luminosities at the 14 TeV LHC,  as a function
of the parton center of mass energy, from 
the CT10, MSTW, and NNPDF2.3 NNLO PDF sets:  Left: $q\bar q$ luminosity; 
Right: $gg$ luminosity, from \cite{QCDworking}}
 \label{fig:pdfs} \end{figure} 

%%%%%%%%%%%%%%%%%%%%%%%%%%%%%%%%%%%%%%%%%%

Still, there are gaps in our knowledge, especially in the components relevant to the study of physics beyond the SM.  The leading PDF distributions disagree in their estimates of the gluon-gluon luminosity function at the mass of the Higgs boson; this accounts for an 8\% systematic uncertainty in the extraction of the cross section for Higgs production.  We have noted already in Section~\ref{Wmass}  that, at the LHC as opposed to 
the Tevatron, the less-constrained antiquark distributions in the proton play a more important role.  Finally,
all parton luminosities are poorly constrained by data for parton-parton invariant masses greater than about 500~GeV. This is clearly shown in Fig.~\ref{fig:pdfs}.

We expect that these difficulties can be addressed using future data from the LHC.  PDFs at the few-percent level of accuracy require theoretical calculations at the NNLO  level.  These are already available for the Drell-Yan process~\cite{DYNNLO}.  The NNLO computation of the total cross section for top quark pair production has recently been completed~\cite{totaltop} and is now being extended to the rapidity distribution. NNLO calculations for 2-jet production are in progress.  Over the next few years, these calculations will be used in conjunction with a very high-statistics data set on jet, top quark, and lepton pair production from the LHC.  The LHCb experiment has an important role to play in the measurement of  rapidity distributions  at high values of rapidity greater than 2.5~\cite{LHCbDY}.

Further improvements in PDFs can result from the program of the LHeC. The data expected will reduce the error in the gluon luminosity to a few percent at the Higgs boson mass and to 5-10\% in the multi-TeV region.

\subsection{QCD coupling constant $\alpha_s$}\label{sec:alphas}

The strength of the QCD coupling is determined by the value of the coupling constant $\alpha_s$, usually expressed as its \MSbar\ value at the $Z$ mass.   The value of this quantity currently quoted by the
 Particle Data Group~\cite{pdg} has  0.6\% uncertainty.  
 We have pointed out in Section~\ref{Hcoupl} that this is a limiting 
systematic error in the evaluation of the SM predictions for Higgs boson couplings.

There are three strategies to improve the precision of $\alpha_s$.  First, $\alpha_s$ can be obtained 
from the $Z$ boson width or the ratio of hadronic to leptonic $Z$ decays.  This technique
 is theoretically unambiguous but currently  is limited by statistics. 
 The present value of $\alpha_s$ from $Z$ decays has  a 4\% uncertainty.  The Giga-Z program at the ILC discussed in Section~\ref{Wmass} should collect 100 times more data and so improve this uncertainty  to 0.4\%.  The very-high-luminosity $Z$ program envisioned for TLEP could decrease this uncertainty further to 0.1\%.  We judge that higher-statistics measurements of $\ee$ event shapes are not competitive with these improvements.

Proposed improvements of PDFs from LHeC will lead to an improved value of $\alpha_s$ from the measurement of PDF evolution.   The expected statistical error would be $\pm$0.2\% from LHeC alone and $\pm$0.1\% from the combination of LHeC and HERA.   The theoretical systematic error for this method is not as well understood, but we estimate this at $\pm$0.5\% once one further order in QCD perturbation theory (N$^3$LO)  is calculated.

The most accurate current values of $\alpha_s$ come from lattice gauge theory.  Higher-statistics lattice estimates and calculation of additional terms in lattice perturbation theory should decrease the current uncertainties over the next decade to 0.3\%.   These improvements will come together with improvements in the values of the quark masses, as discussed in \ref{Hcoupl}.

\subsection{Electroweak corrections to hadron collider processes}

The quest for few-percent accuracy in predictions for hadron colliders brings new elements into play.  In particular, it requires that QED and electroweak corrections  be included in all predictions for LHC.

Three elements are needed here.  Electroweak corrections at NLO order are generally comparable to NNLO QCD corrections.  Thus, calculations intended to be accurate  to NNLO should also 
include electroweak corrections
of order $\alpha_w$ and, if possible, the mixed corrections of order $\alpha_w\alpha_s$.   Electromagnetic corrections to hadronic reactions cannot be consistently included without a set of PDFs derived from formulae that include NLO QED corrections.  This requires a nontrivial modification of PDF fitting programs in order to introduce a photon PDF for the proton. Photon-induced reactions can contribute to LHC processes at the few-percent level, increasing to the 10\% level at higher $pp$ energies.

Finally, at energies of a TeV and above, electroweak Sudakov effects --- negative corrections to two-particle production proportional to $\alpha_w\log^2{s/M_W^2}$ --- can become important.  These are 10\% corrections for Drell-Yan processes producing 3~TeV dilepton systems.  At $pp$ colliders of energy 33 TeV and above, these double logarithmic corrections must be resummed systematically.

\subsection{High-precision calculation}

In the past decade, a revolution in calculational technique has made it possible to derive formulae at NLO for the QCD cross sections for complex multiparton processes such as $pp\to W + 4$ jets and $pp\to t\bar t$ + 2 jets.   This has reduced the size of the theoretical errors in these cross sections from order 1 to 10-20\%.  Methods are now being developed to evaluate general two-parton processes and even some three-parton production processes to NNLO, to reduce these theoretical errors to the few-percent level.

We have already made reference to NNLO calculations of $t\bar t$ and two-jet production.   A very important target here is the cross section for Higgs boson production in association with one or more jets.   Many Higgs measurements at the LHC include jet vetoes to control background from $t\bar t$ production and other sources, so explicit accounting for emitted jets is necessary.  These cross sections often require terms to NNLO for stable summation of the perturbation series.

Beyond the fixed-order perturbation theory, many other aspects of higher-order computation remain to be understood.  NNLO computations often display large logarithms, which should be systematically resummed. The merging of Monte Carlo event generators with NLO QCD calculations is incompletely understood, and new difficulties arise at NNLO.   We are optimistic that Higgs boson production and other QCD processes can be computed to few-percent accuracy, but many challenges remain.

% \subsection{The Message} \label{qcdm}

% The conclusions of the QCD working group can be summarized as follows: \begin{enumerate} \item  Improvements in PDF uncertainties are required. There are   strategies at LHC for these improvements. QED and electroweak corrections must be included in PDFs and in perturbative calculations. \item  An uncertainty is $\alpha_s$ of order  0.1\% may be achievable   through improvements in lattice gauge theory and precision experiments. \item  Advances in all collider experiments, especially on the Higgs   boson, require continued advances in perturbative QCD. \end{enumerate} Experiments on QCD give information on the Particle Physics  Questions \# 1, 2, 8, 9 listed in the Snowmass Summary~\cite{Summary}.
% % %

% \line(1,0){4.3in}
% \vspace{-12pt}
 \section{Fully Understanding the Top Quark} \label{sec:top} %
% \vspace{-12pt}\line(1,0){4.3in}
% %
% \vspace{12pt}

%

The top quark is the heaviest quark and, indeed, the heaviest elementary particle known today.   Its large mass gives it the strongest coupling to the Higgs boson and to other possible particles of the Higgs sector.   The mass of the top quark seems to be anomalously large---though it is sometime argued that it is the masses of all other quarks and leptons that are anomalously small. For all of these reasons, the top quark merits thorough experimental  investigation.

The Tevatron experiments that discovered the top quark produced, in all, about $100,000$ of these particles.  The LHC experiments have already produced 10 million and aim for many billions of top quarks by the end of the HL-LHC.  Future lepton colliders will bring new precision tools to the study of the top quark.   In this section, we will discuss what can be learned from these observations.   More details on all of these topics can be found in the working group report \cite{topworking}.

\subsection{Top quark mass}

Like $\alpha_s$ discussed in Section~\ref{sec:alphas}, the top quark mass is a crucial input parameter for many SM predictions.   It is already the most accurately known quark mass; a 2~GeV uncertainty on this quantity corresponds to a measurement with 1\% precision.  An accurate top quark mass is needed for precision electroweak fits, with an error of 600~MeV on the top quark mass yielding, for example, an error of 5~MeV in $M_W$.   The top quark mass is also an important input to the question of ultimate vacuum stability in the SM~\cite{vacuumstability}. Figure ~\ref{fig:stable} shows that the  measured masses of the Higgs boson and the top quark seem to place the vacuum of the SM  on the very margin of stability. More precision, especially on $m_t$, might indicate whether this is an accident or a hint concerning the parameters
of the Higgs sector.

%%%%%%%%%%%%%%%%%%%%%%%%%%%%%%%%%%%%%%%%

\begin{figure}[htb] \begin{center} 
\includegraphics[width=0.5\hsize]{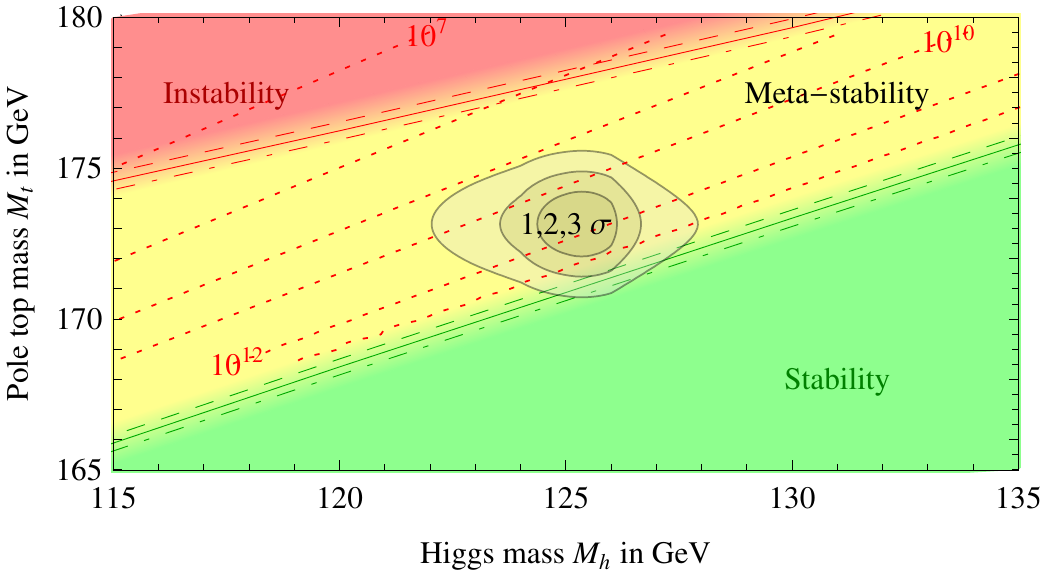} 
\caption{Regions of metastability and instability of the Higgs potential of the SM, as the top quark and Higgs boson masses are varied, from \cite{vacuumstability}.} \label{fig:stable} \end{center} \end{figure} 

%%%%%%%%%%%%%%%%%%%%%%%%%%%%%%%%%%%%%%%%%%

The top quark mass is most precisely defined as an \MSbar\ quantity, evaluated most conveniently at the \MSbar\ top quark mass value itself.  However, experimental determinations of the top quark mass are typically done by kinematic fitting to templates based on leading-order QCD and containing hadronization effects whose uncertainties are poorly controlled.  Thus, the precision determination of the top quark mass requires, first, defining an observable that can be explicitly related to the
  \MSbar\ top quark mass, and, then, measuring that quantity accurately.

One solution to this challenge at the LHC is an idea from CMS~\cite{CMSbell}  to measure the top quark mass from the endpoint of the distribution of the mass of the jet plus lepton system, $m(b\ell)$, that is the
observable product of a semileptonic top quark decay.  In top quark pair production events, $\ell$ is identified as an isolated lepton and the jet is specified as a $b$-tagged jet defined by the
 anti-$k_T$ or a  jet clustering algorithm.
With these definitions,  the distribution of $m(b\ell)$ can be computed in QCD perturbation theory in terms of the perturbative pole mass, which can be related to the  \MSbar\ mass with an error of the order of 200~MeV. The results are largely 
 insensitive to new physics contributions to top quark production.  The endpoint feature in the 
distribution is sharp and strongly dependent on $m_t$.   We expect that this method can reach a total uncertainty of 500~MeV with the statistics of the HL-LHC.   Other methods for measuring the top quark mass at the LHC are discussed in \cite{topworking}.

At lepton colliders, the cross section for top quark pair production near the  threshold has a distinctive rise sensitive to the position of the lowest (unstable) $t\bar t$ bound state.  Extensive theoretical work has evaluated this cross section to NNLO, with resummation of all large logarithms.  The threshold position can be measured to 35~MeV at the ILC and somewhat better at TLEP and muon colliders.  The conversion to the \MSbar\ mass gives a total uncertainty of about 100~MeV. This very accurate value of $m_t$ is well matched to the precision electroweak programs at lepton colliders described in Section~\ref{sec:EW}.

\subsection{Strong and electroweak couplings}

The production and dynamics of top quark pairs at colliders offers many opportunities to test the strong and electroweak couplings of these particles.  At hadron colliders, the dominant pair production mechanism is through QCD.  The current agreement between the predicted and measured values verifies that the absolute strength of the QCD coupling to the top quark is equal to the value of $\alpha_s$ measured for 
light quarks and gluons to about 3\% accuracy.

Changes in the form of the top quark coupling to gluons might be induced by new resonances associated with top quark compositeness.  Possible magnetic or electric dipole couplings can be probed from the kinematics of top final states to better than 1\% at the LHC with 300~\ifb.  Though it is difficult to measure the absolute size of the top quark width at a hadron collider, the $W$ boson helicity fractions in top quark decay are sensitive to modifications of the top quark coupling to the $W$ boson, with similar sensitivity.   The cross section for  single top quark production provides a measure of the weak interaction mixing matrix element $V_{tb}$, which should reach an accuracy of 2.5\% at 300~\ifb.   Couplings of the top quark to the photon and $Z$ boson are constrained by measurements of radiation from a $t\bar t$ state. The HL-LHC is expected to reach sensitivities of a few percent for the photon couplings and 15-20\% for the $Z$ boson couplings.

At lepton colliders, $t\bar t$ pairs are produced through virtual photons and $Z$ bosons, with large interference effects that depend on the beam polarization.  The ILC and CLIC, which can take advantage of large beam polarization, expect to reach sensitivities below the 1\% level for both photon and $Z$ couplings.  Randall-Sundrum models \cite{RS} and other models with top quark and 
 Higgs compositeness predict shifts of the $Z$ boson couplings to $t\bar t$ at the few percent level; these effects could potentially be discovered in the linear collider programs.

\subsection{Rare decays}

The large samples of top quarks available at the LHC allow deep searches for flavor-changing top quark decays.  Neutral current decays of the top quark such as $t \to \gamma c$ or $t \to g c$ are utterly negligible in the SM, with predicted branching ratios smaller than 10$^{-12}$.  These decays can appear with branching ratios as large as $10^{-4}$  in models with an extended Higgs sector or R-parity-violating supersymmetric couplings that bring in two structures of flavor mixing.  Searches for these decays at the HL-LHC can be sensitive to  branching ratios below 10$^{-5}$.

Lepton colliders can also access these flavor-changing couplings in single top quark production, for example, through $\gamma^* , Z^*\to t\bar c, t\bar u$.   Searches for these processes can reach sensitivities close to $10^{-4}$ even in experiments at 250~GeV, below the $t\bar t $ threshold, and below $10^{-5}$ in the full ILC program at 500~GeV.

More details on the specific estimates for each possible neutral current coupling can be found in \cite{topworking}.

\subsection{Searches for new particles related to the top quark} \label{top:newp}

The motivation that we have given for new particles at the TeV scale in Section~\ref{sec:Natural} directly implies the presence of exotic partners of the top quark.   Examples of these particles are top squarks, in models of SUSY, and Kaluza-Klein excitations of top quarks, in models with extra space dimensions.  Searches for these particles have been a very high priority in the LHC program and will continue to be pursued intensively as more data accumulate.

 Searches are designed individually for each type of exotic particle.  The most powerful searches make use of the fact that top quarks resulting from the decay of the partner particle have different polarizations from
 those typically produced in SM pair-production.  This is reflected in the kinematic distributions of the $t\bar t$ final states.    For particles with masses of 1~TeV and above, the preferred method for identifying final-state top quarks is as single  jets with high jet mass and a three-jet substructure~\cite{boosted}.  This ``boosted top'' identification is quite  insensitive to the pileup associated with high luminosity.

Top squarks in SUSY  might be as heavy as other supersymmetric particles.  However, 
 the naturalness arguments we have given in Section~\ref{sec:Natural} indicate that they might, alternatively,
be among the lightest supersymmetric partners.  The LHC experiments have searched extensively for direct pair-production of top squarks that decay to the lightest supersymmetric particle $\widetilde\chi^0$ through $\widetilde t \to t \widetilde\chi^0$ and $\widetilde t \to b \widetilde \chi^+$.   Current searches exclude a top squark up to about 650~GeV in the limit of light electroweak superpartners.  The sensitivity should advance to  about 1.0~TeV at 14~TeV and 300~\ifb, and to 1.2~TeV with 3000~\ifb.

In models with extra space dimensions and models with composite Higgs bosons and top quarks, the expected partners of the top quark are fermions with vectorlike couplings. The searches for these particles are similar to those for hypothetical fourth-generation quarks, but they involve more complex decay patterns, with $T \to Wb$, $T\to tZ$ and $T\to th$.   Searches for these particles that are comprehensive with respect to the decay mode currently exclude vectorlike top partners up to masses  of about 800 GeV.  The 14 TeV stages of the LHC will be able to discover these particles at masses of sensitivity to 1.2~TeV for 300~\ifb\ and to 1.45~TeV for 3000~\ifb.

Models with composite Higgs bosons and top quarks also typically include resonances in the multi-TeV mass region that decay preferentially to $t\bar t$. Randall-Sundrum models, for example, predict a resonance at a mass of a few TeV decaying with high top quark polarization to $t_R\bar t_L$.  The boosted top quark identification described above was developed for the problem of discovering such states and is indeed expected to be very effective.   Applying the same methods to larger data sets, we expect a sensitivity to such resonances up to 4.5~TeV for the 14 TeV LHC with 300~\ifb\ and up to 6.5~TeV with 3000~\ifb.

Additional examples of new particle searches involving top quarks are described in \cite{topworking}.

% \subsection{The Message} \label{topm}

% The conclusions of the Top Quark working group can be summarized as follows: 
% \begin{enumerate} 
% \item The top quark is intimately tied to the problems of electroweak  symmetry   breaking and flavor. 
% \item  Precise and theoretically well-understood measurements of top   quark masses are possible both at LHC and at $\ee$ colliders, in   each case, matching the needs of the precision electroweak program. 
% \item  New top couplings and new particles decaying to top play a key   role in models of electroweak symmetry breaking.  LHC will search for the   new particles directly.  Linear collider experiments will be   sensitive to predicted deviations from the SM in the top quark   couplings.
%  \end{enumerate} Experiments on the top quark give information on the Particle Physics  Questions \# 1, 2, 4, 8, 9 listed in the Snowmass Summary~\cite{Summary}.
% % %

% \line(1,0){\textwidth}
% \vspace{-12pt}
\section{The Path Beyond the Standard Model - New Particles, Forces, and Dimensions} \label{sec:NP} %
% \vspace{-12pt}\line(1,0){\textwidth}
% %
% \vspace{12pt}

%

Models of new physics associated with the TeV mass scale contain a wide variety of new particles.   These include the particles of an extended Higgs sector discussed in Section~\ref{Haddl} and 
partners of the top quark discussed in Section~\ref{top:newp}.   Many of the schemes discussed in Section~\ref{sec:EF-TeV} for explaining Higgs condensation are based on far-reaching principles that require a spectroscopy of new particles containing heavy partners for all SM particles. This includes additional strongly interacting particles, particles with only electroweak interactions, and new vector bosons.  Some of these particles may have lifetimes long enough that their decays are not prompt in a collider experiment. One or more of these particles could be constituents of the cosmic dark matter.

New particles could also introduce new flavor-changing interactions.  Observation of such interactions
of new particles can complement
searches for flavor and CP violation in rare processes.  This subject will be discussed in Section~\ref{sec:flavor}.

The LHC experiments have been able to search for new particles very robustly using a broad range of 
techniques.  In this section, we will discuss how higher energies and luminosities at hadron colliders and new capabilities of lepton colliders will extend these searches.

Models of  Higgs condensation and other TeV-scale phenomena based on the various underlying principles
discussed in Section~\ref{sec:Natural} make qualitatively different predictions for the quantum numbers and mass relations of the  new particle spectrum.    Thus, the first discovery of a new particle beyond the Standard Model will define a direction for an extensive research program, one that will be carried out over decades with multiple complementary experiments.   In this section, we will emphasize the comparative reach of proposed collider programs to make this first discovery. Examples of the consequences of such a discovery will be given in Section~\ref{sec:disc-stor}.   We will have room to discuss only a limited number of examples.  The full range of searches for new particles accessible to TeV energy experiments is described in the  New Particles and Forces  working group report  \cite{NPworking}.

The dependence of search reach on luminosity deserves comment. Away from kinematic limits for a given collider energy, parton-parton luminosity functions scale such that increasing the parton-parton 
center-of-mass energy by a factor 2 decreases the luminosity by a factor of 10.   This rule, which implies that a factor of 10 in luminosity increases the  search reach by a factor of 2 in mass, must break down at masses near the kinematic limit.  At the 14~TeV LHC, the reach increase falls off from the canonical factor of 2  for pair-produced particles with masses well above 1~TeV.

\subsection{New Vector Bosons}

Some predicted particles would show up in collider experiments as distinct resonances.  An example is a color-singlet vector boson associated with an extension of the Yang-Mills symmetry group beyond that of the SM.  Such bosons are required in many contexts, including models with left-right symmetric weak interactions at high energy, models of the Higgsino mass in SUSY, and models with extra dimensions. Models of Higgs composite structure often require breaking of a larger gauge group to the SM symmetry group.

Searches for vector bosons are conducted at hadron colliders by looking for narrow dilepton resonances.  A typical benchmark is sensitivity to the ``sequential SM'' $Z'$, a boson with the couplings of the $Z$ but with a higher mass.  Current results from the LHC require the mass of such a particle to be above 2.5~TeV.  With 14 TeV, it will be possible to discover such a resonance at
4.5~TeV  for 300~\ifb, and at 7~TeV  for 3000~\ifb.  The values of the production cross section and the leptonic forward-backward asymmetry (with respect to the direction of production) give information on the couplings of the $Z'$.   At higher $pp$ energies, the discovery reach increases to 12~TeV for 33~TeV and 30~TeV for the 100~TeV VLHC.

Lepton colliders are sensitive to new vector bosons that interfere with the $s$-channel virtual photon and $Z$ boson in two-fermion production $\ell^+\ell^-\to f\bar f$.  The reach for discovery of a sequential $Z'$ at the ILC at 500~GeV is also about 7~TeV and scales proportional to the center of mass energy for higher energy colliders. Measurements of the $Z'$ signal with two possible beam polarizations and with individual lepton and quark final states gives a large amount of information that may help identify the quantum numbers of the new boson.

\subsection{Supersymmetry}

Searches for supersymmetry (SUSY) encompass a wide range of strategies aimed at different  particles of the SUSY spectrum.

The most generic searches assume that supersymmetric partners of the SM particle carry a conserved quantum number, called R-parity. If the lightest supersymmetric particle is neutral, it will typically be  weakly interacting and will not be observed in a collider detector. Events are then characterized as containing several hadronic jets, associated with decay to the lightest particle plus missing transverse momentum.  In the 
following discussion, we will call these ``jets+$\ETmiss$'' events.  No significant excess of such events
has yet been observed.  The results of the searches are then parametrized by limits on the gluino mass and on a squark mass, assumed common to all squark flavors.   Current LHC results exclude jets+$\ETmiss$ events up to gluino masses of 1.0 TeV and, independently, up to squark masses of
1.3~TeV.   For the future stages of the LHC, we expect to be able to discover such events up to gluino masses of 1.9~TeV and squark masses of 2.3~TeV with 300~\ifb, and to 2.3~TeV and 2.7~TeV with 3000~\ifb. This reflects more than a factor of 2 in increased search power at the 300~\ifb\ stage, and another 20\% with the additional factor of 10 in luminosity.   The gluino discovery reach increases to 4.8~TeV for a 33~TeV $pp$ collider and to 10.2~TeV for the VLHC.

It is possible that the first signal of SUSY would not be given by the generic search just described, but would require a more specialized analysis. Special search techniques are needed in  models  in which mass gaps in the SUSY spectrum are relatively small  (``compressed spectrum''), so that hard jets are not emitted in particle decays, and models in which only the partners of top quarks, or perhaps only color-singlet  supersymmetric particles, are produced at accessible energies.  Strong theoretical motivations have been given for models
of this type.  A compressed spectrum is  needed in many models of supersymmetric dark matter particles  to allow ``coannihilation'' to produce the correct dark matter density~\cite{coann}.  They are also needed in models
in which only the specific particles restricted by the naturalness bounds in Section~\ref{sec:Natural} lie below 2~TeV.   In such models, the first signal of SUSY would come from direct top squark pair production or 
gluino pair production with decay to heavy flavor.  Reach estimates for top squark pair production were given above in Section~\ref{top:newp}.

Models in which SUSY discovery is more difficult at the 8~TeV LHC thus benefit more from the increase in luminosity provided by the  HL-LHC program.  Models for which the first signal of SUSY would be the partners of $W$ and $Z$ bosons can be searched for at the 14 TeV LHC, with discovery expected up to masses of about 500~GeV.  The factor of 10 luminosity increase to HL-LHC increases the reach by a factor of 2 in the analyses~\cite{ATLAS,Xerxes}. 

%%%%%%%%%%%%%%%%%%%%%%%%%%%%%%%%%%%%%%%%% 

\begin{figure}[tb] \begin{center} 
\includegraphics[width=1\hsize]{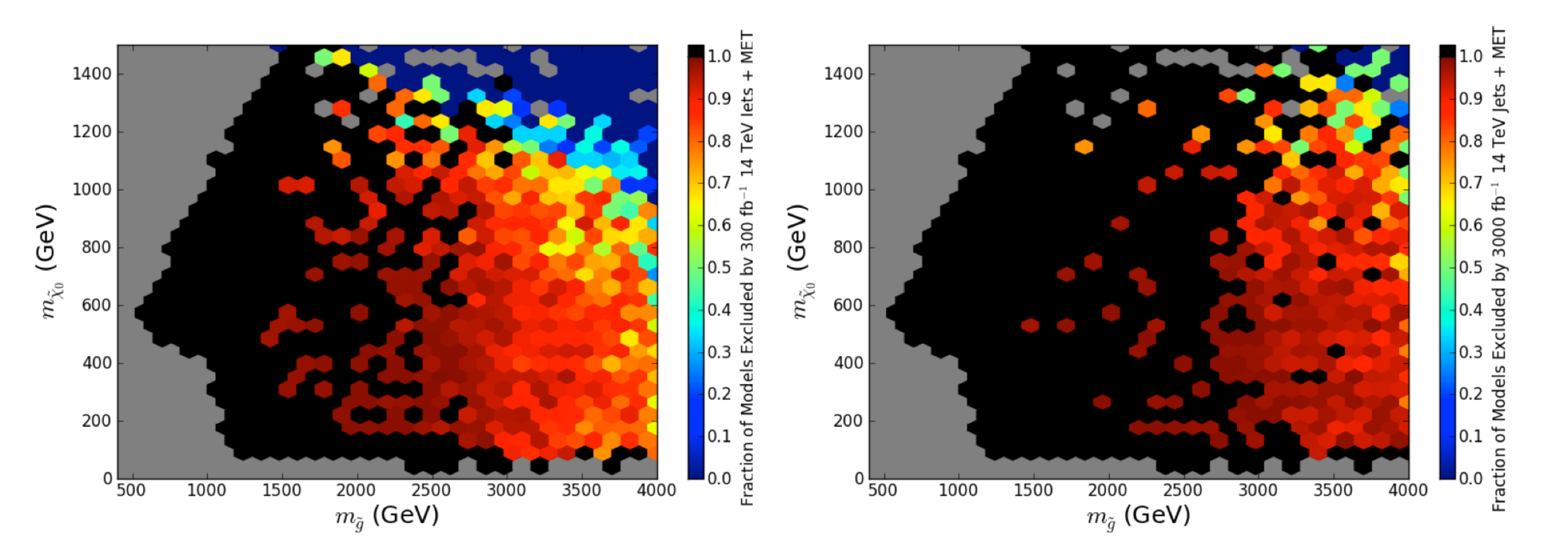}
 \caption{Projections for coverage of the pMSSM 19-parameter model space in searches at the LHC at 300~\ifb\ (left) and 
3000~\ifb\ (right), shown in the plane of gluino mass versus lightest particle mass.  The pixel colors
show the fraction of pMSSM models excluded at that stage of the LHC program:  Black is 
complete exclusion; green is exclusion of 50\% of the models with the given mass values.
  From \cite{pMSSMplot}.}
\label{fig:pMSSMfill} 
\end{center} 
\end{figure} 

%%%%%%%%%%%%%%%%%%%%%%%%%%%%%%%%%%%%%%%%%%

Another way to look at this issue is shown in Fig.~\ref{fig:pMSSMfill}.   The figures show a survey of a large number of SUSY models~\cite{pMSSMplot}  plotted in the plane of gluino mass versus lightest superparticle mass.  The color-coding gives the fraction of models excluded by LHC searches at 14~TeV, with 300~\ifb\ on the left, and with 3000~\ifb\ on the right.  The figures
show that the boundary at which most models are excluded shifts to the right by about 30\%,
but also that a large number of exceptional cases to the left of this boundary are addressed
at higher luminosity.

One more exception should be noted.  The supersymmetric partners of the Higgs boson automatically have small mass splitting of a few GeV. The direct pair-production of these particles  through electroweak interactions is essentially invisible at the LHC, except through searches for initial-state radiation plus invisible particles, described in Section \ref{np:dark}. In a scenario in which these are the lightest supersymmetric particles,  the ability of lepton colliders to be sensitive to very small energy depositions in decay would be crucial to observe and study these particles.  Studies of Higgsino pair production at the ILC are described in \cite{List}.

\subsection{Long-lived particles}

The searches we have described so far assume that all new particles decay promptly at the 
$pp$ collision point.  However, there are many models that give exceptions to this.  ATLAS and CMS have carried out dedicated searches for tracks associated with long-lived massive particles and for particles decaying in the detector, perhaps out of synchronization with the bunch crossings.  Current limits are stronger than those in searches for promptly decaying particles. For example, ATLAS places limits of 310~GeV on a tau slepton, 600 GeV on a top squark, and 985~GeV on a gluino. Should such long-lived particles exist, the LHC detectors trap a sample of them for detailed studies of their decay modes and lifetimes.

\subsection{Dark matter} \label{np:dark}

The search for jets+$\ETmiss$ events discussed above for SUSY applies to a broader class of theories. In Section~\ref{sec:dark}, we introduced WIMP dark matter as a general motivation for new particles with TeV scale masses.   Any model with a new TeV spectroscopy characterized by a new  quantum 
number can  give rise to a dark matter candidate particle.  The requirements are that the lightest new
particle is neutral and that the quantum number is conserved sufficiently that this particle is stable
over the age of the universe. Heavier states carrying QCD color will decay to this lightest particle, producing events with jets and missing transverse momentum. If the partners of quarks are fermionic rather than bosonic, the production cross section will be higher.   The generic SUSY search described above can then be interpreted  as a robust search for models predicting a TeV spectroscopy and dark matter.

It is possible that the heavier states of the TeV spectrum are out of reach kinematically.  Then the discovery of dark matter would require observing direct pair-production of the essentially invisible dark matter particles.  This can be done using the fact that the production of particles in a hard scattering process sometimes also produces gluons or photons radiated from the initial-state particles.  If the particles mediating the pair-production reaction are heavy enough, this initial-state radiation can be at higher momentum than, and
therefore distinguishable from, the ordinary particle production in typical collisions.

%%%%%%%%%%%%%%%%%%%%%%%%%%%%%%%%%%%%%%%%% 

\begin{figure}[tb] 
\begin{center} 
\includegraphics[width=0.5\hsize]{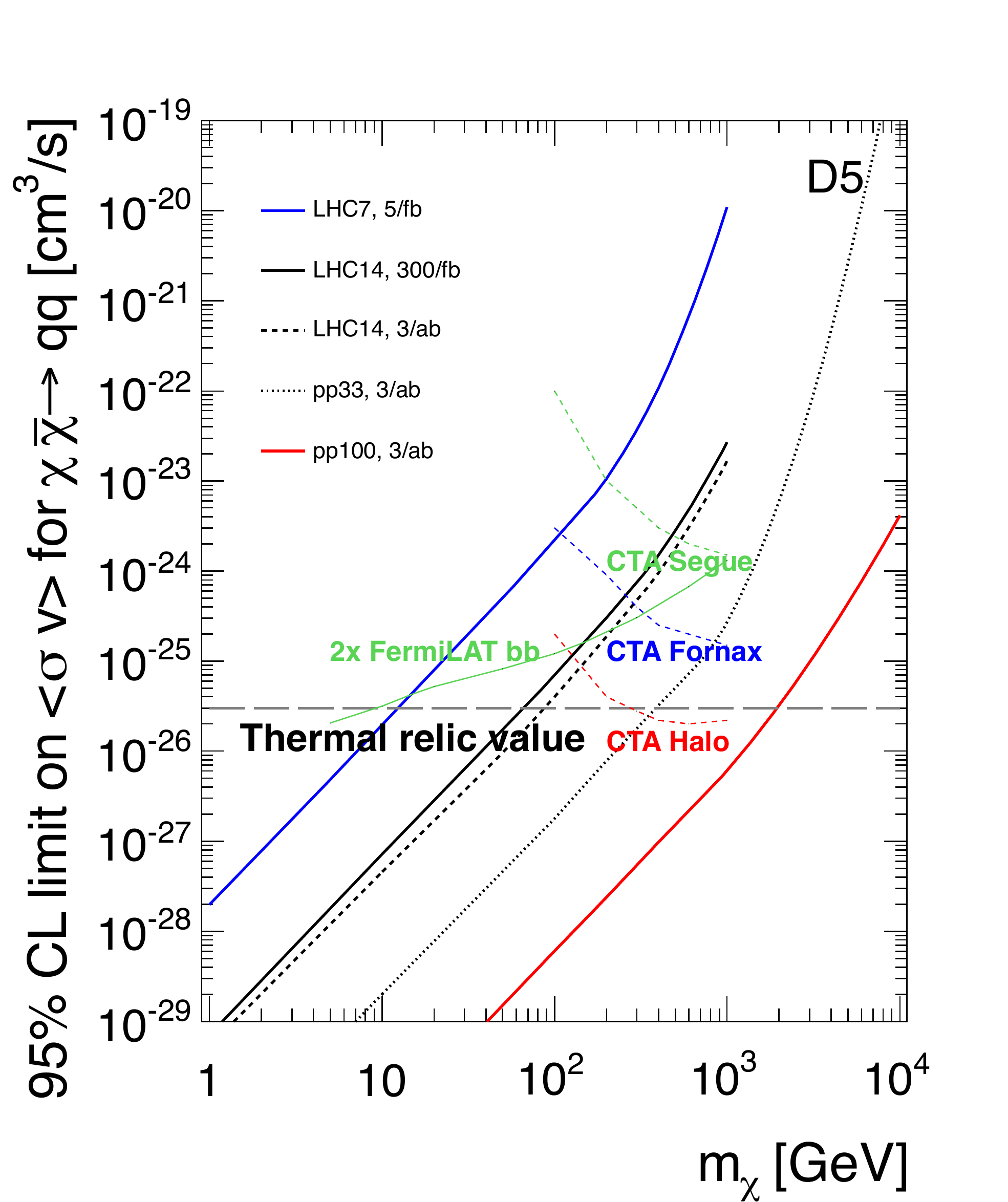} 
\caption{Comparison of 95\% confidence limits on the WIMP pair annihilation cross section in models of 
dark matter, analyzed in an effective operator formalism, for current and proposed facilities. 
 The collider limits are constraints on the 
operator coefficient from searches for missing energy events. The limits from gamma ray 
telescopes are constraints from searches for dark matter annihilation in galaxies of the
local group.  From \cite{Zhou}. }
 \label{fig:DarkMatter} \end{center} \end{figure} 

%%%%%%%%%%%%%%%%%%%%%%%%%%%%%%%%%%%%%%%%%%

Reach estimates for the discovery of dark matter pair-production have been studied systematically in \cite{Zhou}, using an effective operator formalism to describe the coupling of dark matter to SM particles.  This formalism also allows cross sections measured at colliders to be related to rates for direct and indirect detection of dark matter.  
The paper \cite{Zhou} also included limits on an explicit model with a lighter $Z'$ mediator.  Some results of the
effective field theory analysis, comparing limits on the dark matter-nucleon cross section from LHC and higher energy $pp$ colliders to limits from direct detection, are shown in Fig.~\ref{fig:DarkMatter}.    It is noteworthy that the VLHC can place limits on the dark matter particle mass above 1~TeV, close to the unitarity limit for thermal production of such a particle in the early universe.

% \subsection{The Message} \label{npm}

% The conclusions of the New Particles and Forces working group can be summarized as follows: \begin{enumerate} 
% \item  TeV mass particles are needed in essentially all models of new physics. The search for them is imperative.
% \item  LHC and future colliders will give us impressive capabilities for this study. Future programs target new physics at the all-important TeV scale, as can be seen in Fig.~\ref{insight}. 
% \item  The search for TeV mass particles is integrally connected to searches for dark matter. 
% \end{enumerate} 
% Experiments on new particles and forces give information on the Particle Physics  Questions \# 1, 2, 3, 4, 5, 8, 9 listed in the Snowmass Summary~\cite{Summary}.

% %
% \line(1,0){5.4in}
% \vspace{-12pt}
\section{Flavor Mixing and CP Violation at High Energy} \label{sec:flavor} %
% \vspace{-12pt}\line(1,0){5.4in}
% %
% \vspace{12pt}

%

The explanation of Higgs condensation may or may not give insight into the theory of flavor.   It is a mystery why quarks and leptons have a hierarchical mass spectrum and why the weak interactions are not diagonal in flavor and violate CP.   In the SM, these features are parametrized by the fermion-Higgs Yukawa couplings.   The idea that flavors are distinguished only through terms of the structure of the Yukawa couplings  is called ``Minimal Flavor Violation.''

Models of new physics at the TeV scale introduce a large number of new couplings.  These in principle can be proportional to flavor-violating couplings with completely new structures.   Such structures are not needed to build a model of Higgs condensation.   In fact, many specific models---most notably, SUSY with gauge-mediated breaking~\cite{GMSB}---are flavor-blind. It is possible also that some principle, analogous to the GIM mechanism~\cite{GIM}, requires that 
new flavor couplings are related to the Yukawa couplings. 
 In these cases,  there are no new flavor-changing effects beyond the SM arising from the TeV scale.

However, it is also possible that flavor couplings among new particles have a different pattern.  Such couplings can generate rare flavor-changing weak decays.  They can also  affect the phenomenology of the new particles themselves, requiring new strategies for searches.  In this section, we discuss examples of models of this type.   There are many possibilities, only a few of which can be discussed here.  A more complete catalog is given in the working group report \cite{flavorworking}.

\subsection{SUSY with Flavor-Dependent Soft Masses}

In Section~\ref{sec:NP}, we discussed reach estimates for models of TeV spectroscopy that were either blind to flavor or singled out only the third generation.  More general forms of the new particle spectrum are allowed. 
The most important difficulties with flavor observables come when squarks with the same 
gauge quantum
numbers, {\it e.g.}, the partners of $d_R$ and $s_R$, have different masses.  But there is little difficulty in 
giving the partners of $d_R$ and $d_L$ smaller masses than those of the other squarks.  This weakens the
experimental limits on the lightest squark mass.  In fact, it is possible, with other mechanisms to suppress flavor effects of new physics, to allow only the partner of $c_R$ to be light.  This has been explored in detail for SUSY models~\cite{KribsSUSY,PapucciSUSY}.   The current limit on the charm squark in such models is about 300~GeV.

\subsection{R-parity Violating SUSY}

It is possible that the R-parity conservation law that should keep the lightest SUSY partner stable is violated by new interactions.   These interactions necessarily have a complex structure in flavor. One possible form of the
R-parity violating interaction is
\beq
                \lambda^1_{ijk} U_i D_j D_k  \ , 
\eeq{BaryonnoR}
where $U_i$, $D_i$ are the right-handed quarks or their squark partners and $i = 1,2,3$ is the generation number.
This violates baryon number conservation, allowing a squark to decay to two antiquarks.  The interaction must be antisymmetric
in color, and this requires it also to be antisymmetric in the flavors of the two down-type quarks.  Other
possible R-parity violating interactions have the forms
\beq
       \lambda^2_{ijk}   L_i L_j \bar E_k \ , \quad  \lambda^3_{ia}  L_i H_a
\eeq{LeptonnoR}
where $L_i$, $\bar E_i$ are left-handed leptons or antileptons or their slepton partners and $H_a$ is a Higgs
or Higgsino field.  These interactions violate lepton number conservation. 
 The  coefficients of these operators are usually taken to be small to avoid 
unwanted flavor-changing rare decays. In particular, either the operator \leqn{BaryonnoR} 
or the lepton number violating operators must be highly suppressed to avoid rapid proton decay.

In R-parity violating models with the operator \leqn{BaryonnoR} only, SUSY decay chains typically end with the lightest SUSY particle decaying to jets. These jets must have a nontrivial flavor structure, possibly with a $b$ jet always included.

R-parity violation with the Higgs operator in \leqn{LeptonnoR} 
 can produce neutrino masses through the ``Type III seesaw'':  
A neutrino converts to a Higgsino, which then converts back to 
an antineutrino, possibly of a different flavor.  The difference
 between the quark and lepton flavor mixing patterns is explained 
by the statement that the mass matrices come from unrelated operators.  
In such models, the branching ratios of the Higgsino are related to the
 neutrino mixing angles, and this relation can be confirmed by direct measurement~\cite{Vormwald}. 
 In more general seesaw models of neutrino mass, it is possible that the seesaw scale could be the 
TeV scale, setting up other relations  between TeV mass neutral leptons and the neutrino 
mixing  matrix~\cite{HanNu}.

\subsection{Models with Electroweak Baryogenesis}

To explain the asymmetry between the numbers of baryons and antibaryons in the universe, a 
new source of CP violation is needed beyond that of the CKM phase.  One natural place to look for this is 
in an extended Higgs sector.   A second requirement is that the electroweak phase transition 
be first-order. In this case, the transition takes place via the expansion of  bubbles that contain vacuum with a nonzero Higgs field value into the region with zero Higgs field value.  When the 
bubbles fill all of space, the phase transition is complete.  During the period of bubble 
expansion, top quarks scatter from the bubbles because they have zero mass outside 
but are massive inside where the Higgs condensate is present.   A CP phase in the Higgs sector 
makes this scattering asymmetric between matter and antimatter.  Both criteria can be satisfied in
models with multiple Higgs fields~\cite{MorrisseyEWBG}. Measurements of the Higgs 
self-coupling and of CP violation in Higgs decays, described in Sections~\ref{Hscoupl} 
and \ref{Hspin}, test models of this type.

% \subsection{The Message} \label{flm}

% The conclusions of the Flavor and CP working group can be summarized as follows: \begin{enumerate} \item  TeV mass particles may or may not introduce couplings with new   types of flavor violation.  These possible new couplings affect the search methods for new  particles, in many cases, requiring new strategies. \item The search for new particles is integrally connected to searches   for rare flavor changing decays. \end{enumerate} Experiments on flavor associated with new particles give information on the Particle Physics  Questions \# 1, 2, 3, 4, 5, 8, 9 listed in the Snowmass Summary~\cite{Summary}.

%
%-------------------------------------------------------------------------------
\section{Scientific Cases for Future Colliders}
\label{sec:cases}
%-------------------------------------------------------------------------------
%
In the previous sections, we presented the physics opportunities 
for the next steps in the Energy Frontier in terms of individual 
particles and research areas under study.  It is also interesting to assemble these topics
in terms of the experimental program that each accelerator in the 
list given in Section~\ref{sec:accel} will provide.  Integrating over the 
topics in this way, we see that many of these proposed accelerators
have very substantial physics programs that will explore the TeV 
energy scale across a broad range of measurements.

%\mq{This paragraph seems out of place}
%
%\michael{} {
%The next significant running of the LHC will be in 2015 when the center-of-mass energy will be close to 14 TeV and the luminosity will be $10^{34}$\cms. The increase in energy will significantly increase the search reach by close to a factor of 2 in most channels. The first 14 TeV run will accumulate approximately 100~\ifb and will be followed by the ``Phase 1 upgrades'' to the detectors in anticipation of twice the instantaneous luminosity, beginning around 2019.}

In this section, we present the cases for the various accelerators as if each accelerator stood on its own, with no further physics discoveries between now and the time that it begins operation.  However, one should always keep in mind the possibility of  discoveries would 
open up the study of physics beyond the SM.  We have argued already that the likelihood of the discovery of new particle is very high even for the coming runs of the LHC at 14~TeV up to 300~\ifb.
Such a discovery would need to be followed up by 
further exploration that would benefit from accelerators with complementary capabilities or higher energy.
This might, in the end, be the most important benefit of building the accelerators that come later in the
timeline below.  We will expand on this idea in Section~\ref{sec:disc-stor}.

%} {The most important thing to keep in mind in assessing the potential of a future facility: the rhetoric  in 2013  necessarily assumes that every new accelerator stands on its own, with no new physics from any previous experiment. For many reasons already stated, the prospects for new physics emerging within the LHC timeframe are sound. So the physics cases for a machine in $\sim$2030 should be thought of as describing potential. In fact the most important physics case for an accelerator decades from now is to further study the new states discovered between now and then and to follow the physics hints that those new discoveries create. }

%
%-------------------------------------------------------------------------------
\subsection{LHC in this decade: 300~\ifb}    
\label{sec:EF-LHCnow}
\vspace{-10 pt}
First of all, we emphasize the many opportunities that will be
provided
by the coming run of the LHC at 14~TeV.  Operation of the LHC at 
14~TeV to collect 300~\ifb\  of data  offers a tremendous 
increase in the power of
new particle searches.  The increase is close to a factor 2 in mass in most channels.
Many of these searches, such as  the search for the gluino almost
to 2~TeV and the search for vectorlike top partners above 1~TeV,
access
ranges of the masses that are strongly motivated in models of Higgs 
condensation.

This impressive capability is only one aspect of a broad program that
will be carried out at the LHC over the next ten years.  The LHC with 300~\ifb\ will:
\vspace{-3.5mm}
\begin{enumerate}
\item  Clarify Higgs boson couplings, mass, spin, CP to the 10\% level.
\item  Provide the first measurement of the top quark-Higgs boson coupling.
\item  Measure the $W$ boson mass measurement with a precision below 10 MeV.
\item  Make the first measurements of $VV$ scattering.
\item  Measure the top quark mass in a theoretically and experimentally precise way, to 600 MeV.
\item  Measure top quark couplings to gluons, $Z$ and $W$ bosons, and photons with a
precision potentially sensitive to new physics --- a factor 2--5 better
than today.
\item Search for top squarks and top partners and $t\bar{t}$ resonances predicted in models of composite top, Higgs.
\item Provide data for a new generation of PDFs with improved gluon and antiquark distributions.
\item Measure electroweak cross sections in $pp$  collisions, and provide a well-determined
photon PDF.
\item Extend by a factor of 2 the sensitivity to new particles, including SUSY particles, $Z’$, top partners --- key ingredients for models of the Higgs potential --- and the widest range of other possible TeV-mass particles.
\item Carry out deep  searches for dark matter particles accompanied by initial state radiation.
\end{enumerate}
%
%-------------------------------------------------------------------------------
\subsection{High-Luminosity LHC: 3000~\ifb}
\label{sec:EF-HLLHC}

\vspace{-10 pt}
The second high luminosity running of the LHC will be designed for  instantaneous luminosities of $5\times 10^{34}$~\cms. This running with 3000~\ifb\ of accumulated data truly inaugurates the high-precision electroweak era at LHC with few percent precision for most Higgs boson couplings as well as the 5 MeV threshold in $M_W$ mass determination.  The LHC with 3000~\ifb\ will:
\begin{enumerate}
\item  Begin the precision era in Higgs boson couplings, with sensitivities to 2-10\%, and 
1\% for the ratio of $\gamma\gamma$ and $ZZ$ couplings. 
\item  Measure rare Higgs boson decays, $\mu^+\mu^-$ and $Z\gamma$, with 100 M Higgs bosons.
\item  Provide the first evidence of the Higgs boson self-coupling.
\item  Carry out powerful searches for extended Higgs bosons.
\item  Measure the $W$ boson mass to a precision $\pm$ 5 MeV.
\item  Carry out  precise measurements of $VV$ scattering with access to Higgs sector resonances.
\item  Measure the top quark to a precision of  $\pm$ 500 MeV.
\item  Carry out an intensive search for rare, flavor-changing, top quark couplings, with 10 billion top quarks.
\item  Search for top squarks and partners in models of composite top quarks and Higgs bosons in the expected range of masses.
\item  Improve $q$, $g$, and $\gamma$ PDFs to higher $x$ and $Q^2$.
\item  Provide a  20-40\% increase in mass reach for generic new particle searches,
          which can be as much as a 1 TeV step in mass reach.
\item  Extend by a factor of 2 the mass reach for particles produced by the electroweak interactions.
\item  Follow up an earlier  discovery at LHC --- or in dark matter or flavor searches.
\end{enumerate}

%
%-------------------------------------------------------------------------------
\subsection{ILC, up to 500 GeV}
\label{ilc500}
\vspace{-10 pt}
The ILC would run at 250 GeV, 350 GeV, and 500 GeV, in a program that could begin as early as 
the second half of the next decade.  It would study the properties of the Higgs boson, the top quark, and 
possibly also newly discovered particles, in very fine detail.  The ILC,  up 500~GeV center-of-mass energy,
will:
\begin{enumerate}
\item Study tagged Higgs bosons using the reaction $\ee\to Zh$.  This gives model-independent  Higgs boson width and branching ratio measurements, and direct study of all Higgs decay modes, including invisible and exotic decays.
\item  Measure Higgs boson couplings in a model-independent way with percent-level precision  necessary to
probe for new physics beyond the reach of LHC direct searches.
\item  Study the CP properties of the Higgs boson in fermionic channels (e.g., $\tau^+\tau^-$).
\item Carry out a Giga-Z program for EW precision meaurements, and measure the $W$ boson mass to 4 MeV and beyond.
\item  Improve the measurement of triple vector boson couplings by a factor 10, to a precision below expectations for models with Higgs sector resonances.
\item Provide a theoretically and experimentally precise top quark mass to $\pm$100 MeV.
\item Carry out sub-\% measurement of top couplings to $\gamma$ and $Z$, with precision well below expectations in models of composite top quarks and Higgs bosons.
\item Search for rare top couplings in $\ee\to t\bar c, t\bar u$.
\item  Improve the uncertainty in $\alpha_s$ from the Giga-Z program.
\item Search unambiguously for new particles in LHC blind spots --
Higgsino, stealth stop, compressed spectra, WIMP dark matter.
\end{enumerate}

%
%-------------------------------------------------------------------------------
\subsection{ILC at 1 TeV}
\label{ilc1000}
\vspace{-10 pt}
An extension of ILC to 1~TeV will access additional Higgs boson reactions
for precision study and, possibly, also reach new particle thresholds.  The ILC at 1~TeV will:
\begin{enumerate}
\item Provide a precision measurement  of the Higgs boson coupling to the top quark, to  2\%. 
\item Measure the  Higgs boson self-coupling to 13\%.
\item  Discover  any extended Higgs boson states coupling to the $Z$, up  to 500 GeV.
\item  Improve the  precision of triple gauge boson couplings by a
  factor of 4 over 500 GeV results.
\item  Carry out a model-independent search for new particles 
with coupling to $\gamma$ or $Z$ to 500 GeV.
\item  Search for $Z’$ using $\ee\to f\bar f$  to masses of about 5 TeV, a reach
  comparable to LHC for similar models, and provide multiple observables for diagnostics of the $ Z’$ couplings.
\item Follow up any previous discovery of a new particle, with a
search for electroweak partners, a 1\% precision mass measurement, observation of the complete decay profile, model-independent measurement of cross sections, branching ratios, and couplings with polarization observables, and a search for flavor and CP-violating interactions.
\end{enumerate}

%
%-------------------------------------------------------------------------------
\subsection{CLIC: 350 GeV, 1 TeV, 3 TeV}
\label{clic}
\vspace{-10 pt}
Extremely high energies in $\ee$ collisions will likely 
require technologies beyond that envisioned for the ILC.  CLIC is proposed as
an $\ee$ collider capable of multi-TeV energies.  It
would probe Higgs boson self-couplings and exotic
 scattering of both standard model particles and any new particles 
found or hinted at in earlier machines.  CLIC will:
\begin{enumerate}
\item  Provide a precision measurement of the Higgs boson coupling to top, to  2\%.
\item  Measure the Higgs boson self-coupling to 10\%.
\item  Discover any extended Higgs boson states coupling to the $Z$, up  to 1500 GeV.
\item  Improve the  precision of triple gauge boson couplings by a
  factor of 4 over 500 GeV results.
\item  Make precise measurement of $VV$ scattering, sensitive to Higgs boson  sector
  resonances.
\item Carry out a model-independent search for new particles with coupling to $\gamma$ 
or $Z$ to 1500 GeV, the expected range of masses for electroweakinos and WIMPs.
\item Search for $Z$’ using $e^+ e^- \rightarrow f \bar{f}$ accessing masses above 10 TeV.
\item  Follow up any discovery of new particles with the elements listed
        for the 1~TeV ILC, but accessing higher mass states.
\end{enumerate}

%
%-------------------------------------------------------------------------------
\subsection{Muon Collider: 125 GeV, 
       350 GeV, 1.5 TeV,  3 TeV}
\label{mu}
\vspace{-10 pt}
A muon collider holds promise as a technique for reaching very high energies in lepton-lepton
collisions and for $s$-channel production of the Higgs boson  and possible additional Higgs states.
Studies of the muon collider are not yet mature, particularly in designing a detector that can overcome
the background from decays of the muons circulating in the ring.  However, promising first results on
physics analysis including machine backgrounds were
reported at Snowmass.   A muon collider will:
\begin{enumerate}
\item Provide capabilities similar to those of ILC and CLIC described above.
\item Produce the Higgs boson, and possible heavy Higgs bosons, as $s$-channel resonances This allows a sub-MeV Higgs boson mass measurement and a direct Higgs boson width measurement.

\end{enumerate}

%
%-------------------------------------------------------------------------------
\subsection{Photon Collider}
\label{photon}
\vspace{-10 pt}
Another technique for producing Higgs bosons in the $s$-channel is to convert an
electron collider to a photon collider by backscattering laser light from the electron beams.  This
allows resonance studies at 80\% of the electron center-of-mass energy.  Photon colliders will:
\begin{enumerate}
\item Produce Higgs or extended Higgs bosons as $s$-channel resonances, offering percent-level accuracy in $\gamma\gamma$ coupling.
\item  Study CP mixture and violation in the Higgs sector using polarized photon beams.
\end{enumerate}

%
%-------------------------------------------------------------------------------
\subsection{TLEP, Circular $e^+e^-$}
\label{tlep}
\vspace{-10 pt}

An $\ee$ collider in a very large tunnel offers the possibility of very large integrated luminosity 
samples at 250~GeV and below and reasonable integrated luminosity at 350~GeV, especially if multiple detectors can be used simultaneously.   TLEP will:
\begin{enumerate}
\item Offer the possibility of up to 10 times higher luminosity than linear $\ee$ colliders at 250 GeV, with 
corresponding factor of 3 improvements in  boson couplings measurements.
\item  Carry out precision electroweak measurements 
 that could improve on ILC by a factor of 4 in $\ssteff$, a factor of 4 in $M_W$, and a factor of 10 in $M_Z$.
\item Provide a theoretically and experimentally precise top quark mass to $\pm$100 MeV.
\item  Search for rare top couplings in $\ee\to t\bar c, t\bar u$.
\item Offer a possible improvement in $\alpha_s$  by a factor of 5 over Giga-$Z$, to 0.1\% precision.
\end{enumerate}

%
%-------------------------------------------------------------------------------
\subsection{VHLC, at 100 TeV}
\label{vlhc}
\vspace{-10 pt}
One of the ideas that gained momentum at Snowmass  was renewed interest in a Very Large Hadron Collider (VLHC).  Our study recommends reinvigorating R\&D  toward realization of a VLHC.   A VLHC will:
\begin{enumerate}
\item Provide high rates for double Higgs boson production, and a measurement of Higgs boson self coupling to 8\%.
\item Search sensitively for  new Higgs bosons and states associated with extended Higgs sectors at 1~TeV.
\item Dramatically improve the sensitivity to vector boson scattering and multiple vector boson production.
\item Increase the  search reach for new particles associated with naturalness---including SUSY particles, top partners, and resonances---by almost an order of magnitude in mass over LHC.  This corresponds to two 
orders of magnitude in fine-tuning.
\item Search for WIMP dark matter up to TeV masses, possibly covering the full natural mass range.
\item Follow up any discovery at LHC --- or in dark matter or flavor searches --- with  more detailed measurements, and with  searches for related
higher-mass 
particles. Both luminosity and energy are relevant.
\end{enumerate}

\section{Discovery stories}
\label{sec:disc-stor}

Another way to survey the physics topics presented in
Sections~\ref{sec:higgs}--\ref{sec:flavor}
is to consider the consequences of a discovery of new physics at the
LHC later in this decade.  We have emphasized, first, that the presence
of
new particles at the TeV scale is necessary to build a physics
explanation of electroweak symmetry breaking, and, second, that the 
coming run of the LHC, up to 300~\ifb, will improve the depth of
searches 
for new particles by more than a factor of 2.   The conclusion from these
statements is that the discovery of new physics at the LHC is likely.
This means that we should have a plan for following up this discovery and 
exploring its implications.  This program of course depends on the 
nature of the particle discovered, so a full analysis would be
presented
as a large number of case studies.  

In this section, we give two illustrative examples of specific discoveries that 
might be made at the LHC in the coming decade, and the physics programs that 
would follow from them.
 Further examples of these ``discovery stories'' are 
presented in the working group reports.

\subsection{Well-Tempered Neutralino in SUSY}
\label{gluinoDS}

The New Particles and Forces working group \cite{NPworking} considered in some detail 
the consequences of a particular SUSY model that could be 
discovered at the LHC with 300~\ifb.  This particular model has a
gluino at 1.9~TeV, squarks ranging in mass from 1.3~TeV to 2.6~TeV,
and  bino and Higgsino states near 200~GeV.   The bino and Higgsino
are  assumed to mix so that the lightest supersymmetric particle
(LSP)  would 
be a dark matter particle with the correct thermally generated cosmic
density.  This dark matter scenario, originally developed in \cite{Feng},
 is called the ``well-tempered
neutralino''~\cite{WTN}. 

The LHC at 300~\ifb\ would observe a robust jets+$\ETmiss$ signal.   The signal would be dominated by the decay
of the lighter squarks to a quark jet plus the unobserved LSP.   The
mass difference could be measured from kinematic distributions.
Assuming that the LSP was light, this would also give an estimate of
the
squark mass.  With the measured cross section, this would favor
SUSY over models with fermionic partners.

The HL-LHC would produce some of the heavier squarks and the gluino.
Detailed
kinematic measurements would  identify
at least one more mass scale in the spectrum and give further evidence for the
SUSY hypothesis. 
The direct production of electroweak states would not be observed at
the LHC, because
these states have a compressed spectrum. 

A lepton collider with center-of-mass energy of 500 GeV would be 
able to pair-produce the Higgsinos and observe their decays to the
LSP.  Measurement of the polarized cross sections would give
information
about the quantum numbers of the electroweak states.  It would 
also give an indirect determination of the mass of the electron-type
slepton (750 GeV in this model) to 10 GeV.  Using this information, it would be possible to evaluate the LSP annihilation cross section  and show that it was consistent with that required for a dark matter particle.

Experiments at higher-energy colliders would be needed to discover
the heaviest sleptons (at 3.3 TeV) and squarks (at 2.6 TeV).
Eventually,
the complete SUSY spectrum would be determined, and the 
data on the mass spectrum could be used to deduce the pattern of
SUSY breaking.

\subsection{$t\bar t$ Resonance in a Randall-Sundrum model}
\label{sec:resDS}

An alternative scenario is based on a Randall-Sundrum 
model with top and Higgs compositeness~\cite{RS}.    The first evidence
of this model would be the discovery of a resonance in $pp$
collisions that decayed to $t\bar t$. Such a resonance at 3~TeV
would be discovered at the LHC with 300~\ifb.  Study of kinematic
distributions of the $t\bar t$ final state would reveal that the 
top quarks were highly polarized, a prediction of this model.

The HL-LHC would discover an electroweak singlet top quark partner,
and, possibly also, a doublet of quarks with vectorlike coupling to 
the electroweak interactions.   It is possible that very accurate 
studies of the $t\bar t$ spectrum would also reveal the presence of
a color-singlet resonance, somewhat below 3~TeV.   Its
higher-statistics
study of the TeV resonance might reveal a decay to $t\bar c$ with
branching ratio $10^{-3}$.

A lepton collider at 500 GeV would observe a
significant 3\% enhancement of the right-handed top quark coupling to the
$Z$ boson.  The pattern of Higgs boson couplings would be shifted by the
influence of the 
new top quark partners,  with a substantial increase
in the $t\bar t$
coupling, a 4\% decrease in the  $b\bar b$ coupling, and a 2\% increase 
in the $\gamma\gamma$ coupling.  The last of these effects would be 
 observed by combining the high statistics 
measurement of $BR(\gamma\gamma)/BR(ZZ^*)$ from the HL-LHC with the
precise measurement of the Higgs coupling to $Z$ at a lepton collider.

These measurements would give a tantalizing first glimpse of the
structure
of the underlying composite Higgs model.  Experiments at higher
energy colliders capable of producing resonances up to 20~TeV in 
mass  would be needed to explore the full structure of the spectrum of
states.

\section{Messages from the working groups}\label{sec:Messages}

 In this section, we
collect the  summary statements from each of the six working groups.

\subsection{Higgs: the message}

The conclusions of the Higgs Boson working group can be summarized as follows: 
\begin{enumerate} 
\item Direct measurement of the Higgs boson is the key to   understanding electroweak symmetry breaking.   The fact that the Higgs boson  appears as a light, apparently fundamental, scalar particle needs  explanation. A research program focused on the Higgs couplings to fermions and vector bosons and achieving a precision of a few percent or less is required to address these questions. 
\item  Full exploitation of the LHC is the path to few percent   precision in the Higgs coupling and to a 50 MeV precision in the determination of the Higgs mass. 
\item Full exploitation of a precision electron collider is the path to a model-independent measurement of the Higgs boson width and a sub-percent measurement of the Higgs couplings. Such precision is necessary to 
probe for new physics beyond the reach of LHC direct searches.
 \end{enumerate}

\subsection{Electroweak: the message}
The conclusions of the Electroweak Interactions working group can be summarized as follows: \begin{enumerate} 
\item Precision measurements of the $W$ and $Z$ bosons have the potential to probe indirectly for new particles with TeV masses.  This precision program is within the capabilities of LHC, linear $\ee$ colliders, and TLEP. 
\item  Measurement of vector boson interactions will  probe for new dynamics in the Higgs sector. In such theories, we expect correlated signals in triple and quartic gauge boson couplings.  The LHC and linear colliders will have sensitivity into the mass region above 1 TeV. 
\end{enumerate} 

\subsection{QCD: the message}

The conclusions of the QCD working group can be summarized as follows: 
\begin{enumerate} 
\item  Improvements in PDF uncertainties are required. There are   strategies at LHC for these improvements. QED and electroweak corrections must be included in PDFs and in perturbative calculations. 
\item  An uncertainty in $\alpha_s$ of order  0.1\% may be achievable   through improvements in lattice gauge theory and precision experiments. 
\item  Advances in all collider experiments, especially on the Higgs   boson, require continued advances in perturbative QCD. \end{enumerate} 

\subsection{Top quark: the message}

The conclusions of the Top Quark working group can be summarized as follows: 
\begin{enumerate} 
\item The top quark is intimately tied to the problems of electroweak  symmetry   breaking and flavor. 
\item  Precise and theoretically well-understood measurements of top   quark masses are possible both at LHC and at $\ee$ colliders, matching, for each,  the needs of the precision electroweak program.
\item  New top quark couplings and new particles decaying to top quarks play a key   role in models of electroweak symmetry breaking.  LHC will search for the   new particles directly.  Linear collider experiments will be   sensitive to predicted deviations from the SM in the top quark   couplings.
 \end{enumerate} 

%%%%%%%%%%%%%%%%%%%%%%%%%%%%%%%%%%%%%%%%% 

\begin{figure}[t] \begin{center} 
\includegraphics[width=0.8\hsize]{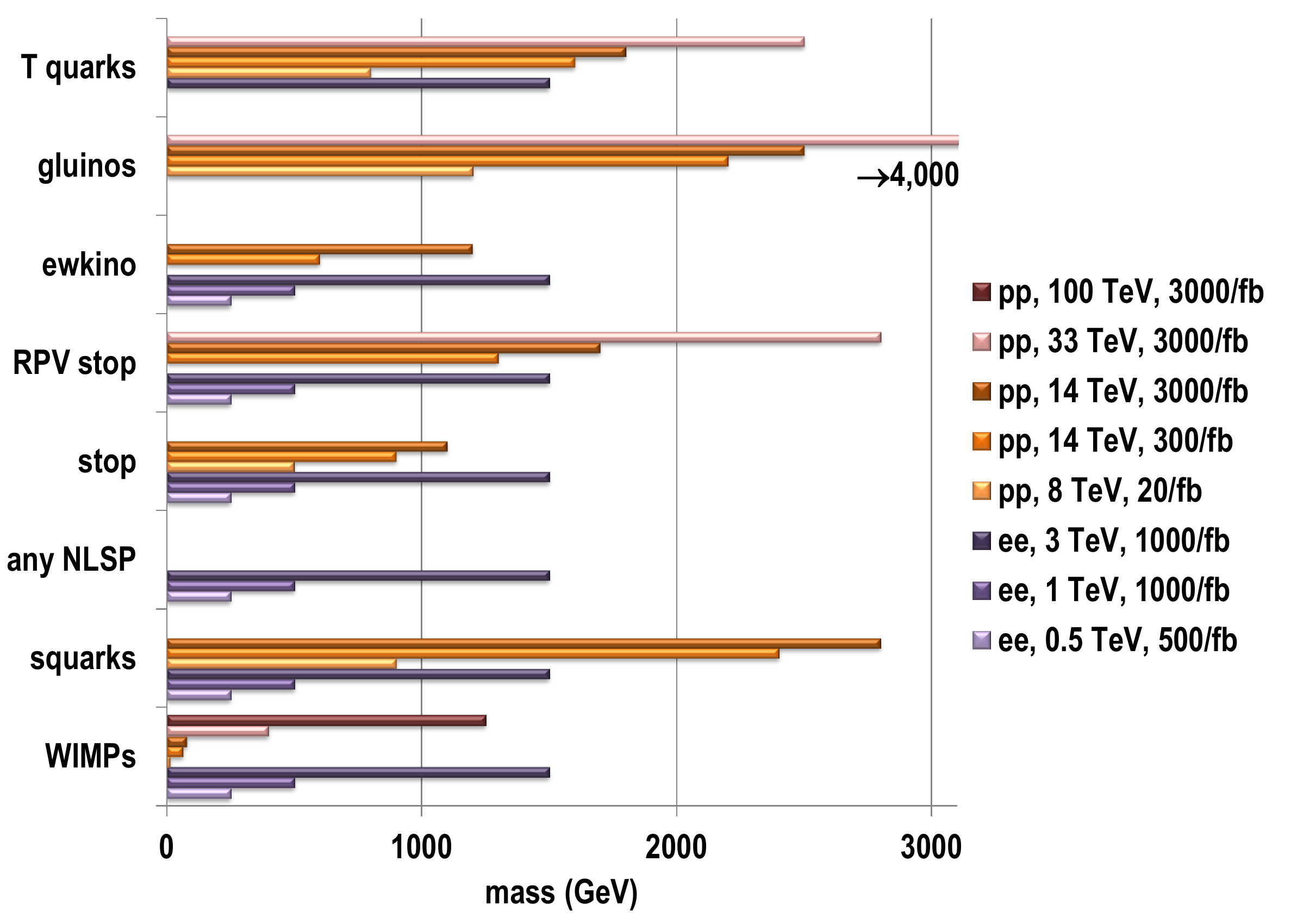} 
\caption{Examples of 95\% confidence upper limits for new particle searches at 
proposed $pp$ 
and $\ee$ colliders.}
\label{insight} 
\end{center} \end{figure} 

%%%%%%%%%%%%%%%%%%%%%%%%%%%%%%%%%%%%%%%%%%

\subsection{New particles and forces: the message}

The conclusions of the New Particles and Forces working group can be summarized as follows: \begin{enumerate} 
\item  TeV mass particles are needed in essentially all models of new physics. The search for them is imperative.
\item  LHC and future colliders will give us impressive capabilities for this study. Future programs target new physics at the all-important TeV scale, as can be seen in Fig.~\ref{insight}. 
\item  The search for TeV mass particles is integrally connected to searches for dark matter. 
\end{enumerate} 

\subsection{Flavor: the message}

The conclusions of the Flavor and CP working group can be summarized as follows: \begin{enumerate} \item  TeV mass particles may or may not introduce couplings with new   types of flavor violation.  These possible new couplings affect the search methods for new  particles, in many cases, requiring new strategies. \item The search for new particles is integrally connected to searches   for rare flavor changing decays. \end{enumerate} 

\section{Conclusions} \label{sec:EF-Concl}

In this report, we have described the future program of research at high-energy colliders and summarized the efforts of six Snowmass 2013 Energy Frontier working groups. The detailed results that we
have reviewed lead to a set of points that deserve special emphasis.

{\bf  The search for new particles with masses of order 1 TeV is a central issue in particle physics.}
 The mysteries associated with the Higgs field and dark matter give compelling arguments for a new
particle spectroscopy at the TeV mass scale.
Such particles must exist to provide a physics explanation for the Higgs condensation and symmetry
breaking.
   We need to know their nature, and their implications for the laws of physics at very short distances.  We have given many examples in which these particles address other major questions of particle physics, including the questions of flavor, dark matter, and the unification of forces.

{\bf   The central capability of high-energy collider experiments is to produce massive
 elementary particles directly.}  In Energy Frontier experiments, we observe the $W$ and $Z$ bosons, the 
top quark, and the Higgs boson as real particles whose production and decay we can study in detail.
The same will be true for any new particles that we can create.  This is a unique, direct, and powerful method
to learn about the laws of physics.

{\bf  There are three essential aspects in the  exploration of the physics of the TeV mass scale:} 
\begin{enumerate}
\item  We must study the Higgs boson itself in as much detail as possible, searching for signs of a larger Higgs sector and the effects of new heavy particles.
\item We must search for the imprint of the Higgs boson  and its possible partners on the couplings of
the $W$ and $Z$ bosons and the top quark.
\item We must search directly for new particles with TeV masses that can address  important problems in fundamental physics.
\end{enumerate}

{\bf  This program can  be realized at accelerators now envisioned to operate in the coming years.}
\begin{enumerate} 
\item We have emphasized the great opportunity that is being provided by the coming operation of the LHC at 14~TeV.   The next stage of the LHC will double the range of searches for new particles and give similar leaps in capability for other probes of TeV physics. 
\item We have projected quantitatively what will be achieved in running the LHC at high luminosity, \ to ultimately acquire 3000~\ifb\ of data per experiment.   For some physics topics, the gain is incremental. For
others, in particular, the precision study of Higgs couplings and the search for new particles with only electroweak interactions, the high luminosity moves us to a qualitatively new level.  Viewed as a whole, 
this is a compelling physics program. 
\item We have listed many essential contributions to the exploration of the TeV scale that can be provided by lepton colliders.  These include precision studies of the Higgs boson, the $W$ and $Z$ bosons, and the top quark, capable in all cases of discovering percent-level corrections to SM predictions expected as the effects of TeV mass particles. The construction and operation of the ILC in Japan will realize these goals.
\item We have emphasized that the quest to understand the TeV scale will not be finished with the results of accelerators of the next generation.   It is likely that the discovery of new particles 
at the next stage of collider physics will open  a definite path for exploration to still higher energies.  Our study called attention, in 
particular, to the capabilities of a VLHC for further exploration of the TeV mass scale. 
The journey to still higher energies begins with renewed effort to bring advanced accelerator technologies to reality.
 
\end{enumerate}

We emphasized in our introduction that the discovery of the Higgs boson changes everything. This discovery points to potentially profound modifications of the laws of physics at energies relatively close to those we now access at accelerators.  The quest for new phenomena and the insights they will provide has just begun.

\Acknowledgements

The Energy Frontier conveners  are profoundly grateful to the 26 dedicated conveners of our working groups: Kaustubh Agashe, Marina Artuso, John Campbell,  Sally Dawson, Robin Erbacher, Cecilia Gerber,
Yuri Gershtein, Andrei Gritsan, 
Kenichi Hatakeyama, Joey Huston, 
Ashutosh Kotwal, 
Heather Logan, Markus Luty, Kirill Melnikov, Meenakshi Narain, Michele Papucci, Frank Petriello, Soeren Prell,  Jianming Qian, Reinhard Schwienhorst,
Chris Tully,
Rick Van Kooten, Doreen Wackeroth, Lian-Tao Wang,  and Daniel Whiteson. 

We are also grateful to the many scientists who assisted us in various ways through the process. We have benefited from those who have given us technical advice, from those who maintained our contact with the major collaborations, and from 
 those who took time to help us with individual tasks.

Our technical team was:  Jeff Berryhill, Sergei Chekanov,  Tom LeCompte, Sanjay Padhi,  Eric Prebys, Tor Raubenheimer, and Eric Torrence.  We also thank Markus Klute and Mark Palmer from the Capabilities group.

Our advisors from the major experiments were: for ATLAS: Ashutosh Kotwal; for CMS: Jim Olsen; for LHCb: Sheldon Stone; for ILD: Graham Wilson; for SiD: Andy White; from CLIC: Mark Thomson; from the Muon Collider: Ron Lipton; and for the VLHC: Dmitri Denisov.

We are very grateful to Ashutosh Kotwal, Laura Reina, Chris Tully, Bob Kehoe, and Peter Onyisi for organizing Energy Frontier workshops at their institutions.  We extend special gratitude to Howard Gordon and Sally Dawson of BNL, Gordon Watts of UW, and Lars Bildsten of the KITP for hosting the major meetings of the study.  We warmly thank Gabriella Sciolla and Ketino Kaadze for organizing the Snowmass@CERN
meetings and study groups.

Finally, we thank our fellow Snowmass 2013 conveners who asked ``tough questions'' and answered them.

\end{document}

%%%%%%%%%%%%%%%%%%%%%%%%%%%%%%%%%%%%%%%%%%%%%%%%%%%%%%%%%%%%%%%%%%%%%%%%% %% %%   use this format to include an .pdf figure into your paper %% 
%\begin{figure}[htb] 
%\begin{center} 
%\includegraphics[width=0.3\hsize]{Magnetism/magnet.pdf} %\caption{Plan of the magnet used in the Mesmeric studies.} %\label{fig:magnet} 
%\end{center} 
%\end{figure} 
%%%%%%%%%%%%%%%%%%%%%%%%%%%%%%%%%%%%%%%%%%%%%%%%%%%%%%%%%%%%%%%%%%%%%%%%%%%

%%%%%%%%%%%%%%%%%%%%%%%%%%%%%%%%%%%%%%%%%%%%%%%%%%%%%%%%%%%%%%%%%%%%%%%%% %% %%   use this format to include a LaTeX table  into your paper %%
\begin{table}[t] 
\begin{center}
\begin{tabular}{l|ccc} 
Patient &  Initial level($\mu$g/cc) &  w. Magnet & w. Magnet and Sound
\\ \hline  
Guglielmo B.  &   0.12     &     0.10      &     0.001  \\  
Ferrando di N. &  0.15     &     0.11      &  $< 0.0005$ \\ \hline 
\end{tabular} 
\caption{Blood cyanide levels for the two patients.} 
\label{tab:blood} 
\end{center} 
\end{table} %%%%%%%%%%%%%%%%%%%%%%%%%%%%%%%%%%%%%%%%%%%%%%%%%%%%%%%%%%%%%%%%%%%%%%%%%%%
%\input HeavyPhotons/wgreport.tex

%%%%%%%%%%%%%%%%%%%%%%%%%%%%%%%%%%%%%%%%%%%%%%%%%%
%%%%%%%%%%%%%%%%%%%%%%%%%%%%%%%%%%%%%%%%%%%%%%%%%%
%%%   Your subdirectory (here Magnetism) should include
%%%    the files:
%%%           wgreport.tex
%%%           authorlist.tex
%%%         and all needed figures in pdf format
%%%%%%%%%%%%%%%%%%%%%%%%%%%%%%%%%%%%%%%%%%%%%%%%%%%%
%%%%%%%%%%%%%%%%%%%%%%%%%%%%%%%%%%%%%%%%%%%%%%%%%%%%


\begin{thebibliography}{99}
\bibitem{LGTforHiggs}
This is reviewed in Section 4 of the QCD working group 
report \cite{QCDworking}.
\end{thebibliography}
\end{document}